\documentclass[9pt,lineno]{elife}
\usepackage{indentfirst}
\usepackage{lipsum} 
\usepackage[version=4]{mhchem}
\usepackage{siunitx}
\DeclareSIUnit\Molar{M}
\DeclareMathOperator*{\argmax}{arg\,max}
\DeclareMathOperator*{\argmin}{arg\,min}
\usepackage[ruled,vlined]{algorithm2e}
\usepackage{subcaption}

\title{Efficient Sampling-Based Bayesian Active Learning for synaptic characterization}

\author[1*]{Camille Gontier}
\author[1]{Simone Carlo Surace}
\author[2,3]{Igor Delvendahl}
\author[2,3]{Martin Müller}
\author[1,4]{Jean-Pascal Pfister}
\affil[1]{Department of Physiology, University of Bern, Bern, Switzerland}
\affil[2]{Department of Molecular Life Sciences, University of Zurich, Zurich, Switzerland}
\affil[3]{Neuroscience Center Zurich, Zurich, Switzerland}
\affil[4]{Institute of Neuroinformatics and Neuroscience Center Zurich, University of Zurich/ETH Zurich, Zurich, Switzerland}

\corr{camille.gontier@unibe.ch}{CG}

\begin{document}

\maketitle

\begin{abstract}
Bayesian Active Learning (BAL) is an efficient framework for learning the parameters of a model, in which input stimuli are selected to maximize the mutual information between the observations and the unknown parameters. However, the applicability of BAL to experiments is limited as it requires performing high-dimensional integrations and optimizations in real time: current methods are either too time consuming, or only applicable to specific models. Here, we propose an Efficient Sampling-Based Bayesian Active Learning (ESB-BAL) framework, which is efficient enough to be used in real-time biological experiments. We apply our method to the problem of estimating the parameters of a chemical synapse from the postsynaptic responses to evoked presynaptic action potentials. Using synthetic data and synaptic whole-cell patch-clamp recordings, we show that our method can  improve the precision of model-based inferences, thereby paving the way towards more systematic and efficient experimental designs in physiology.\end{abstract}

\section{Introduction}

In neuroscience, machine learning, and statistics, a central problem is that of inferring the parameters $\theta$ of a model $\mathcal{M}$. For instance, in supervised learning, one may want to learn the parameters of a Deep Neural Network (DNN) so as to minimize the difference between its output and training labels; in this case, $\mathcal{M}$ represents the DNN to be trained, and $\theta$ represents its weights and biases. Similarly, in biology, the parameters of a system can be studied by fitting a biophysical model to recorded observations.
In most cases, these parameters can be neither directly measured nor analytically computed, but can be inferred using the recorded outputs of the system $y$ as a response to input stimuli $x$. In biology, the physical quantities of a system (e.g. an organ, a cell, or a synapse) can be estimated by deriving a generative biophysical model $\mathcal{M}$ of the system, and by fitting its parameters $\theta$ to the observed responses $y$ to experimental inputs $x$.
By computing the likelihood of the outputs given the inputs and the parameters $p(y | x, \theta)$, it is possible to obtain either a point-based estimate of the parameters such as the maximum likelihood parameters $\theta_{\rm ML}$ or the maximum a posteriori parameters $\theta_{\rm MAP}$ \citep{barri2016quantifying}, or to compute the full posterior distribution $ p(\theta | x, y) \propto p(y | x, \theta)$ using for instance the Metropolis-Hastings (MH) algorithm  \citep{bird2016bayesian}.

However, the accuracy of these estimates critically depends on the pair $(x,y)$, and especially on how the successive input stimuli $x = x_{1:T}$ are chosen. For instance, training a DNN on non independent and identically distributed (i.i.d.) training examples (i.e. blocked training) will lead to catastrophic forgetting \citep{flesch2018comparing}. On the other hand, most experiments in biology still rely on pre-defined and non-adaptive inputs $x_{1:T}$, which may not yield sufficient information about the true parameters of the studied system. Consequently, experiments often require more observations or repetitions to reach a certain result, which increases their cost, time, and need for subjects. 

An efficient framework to alleviate this issue is called Bayesian Active Learning (BAL). Knowing the current estimate of the parameters, the experimental protocol (i.e. the next input $x_{t+1}$) can be optimized on the fly to maximize the mutual information between the recordings and the parameters (Figure \ref{fig:setup_synapse} (b)).
BAL is a branch of Optimal Experiment Design (OED) theory \citep{emery1998optimal,sebastiani2000maximum,ryan2016review}. It has already been used in neuroscience to infer the parameters of a Generalized Linear Model (GLM) \citep{lewi2009sequential}, the nonlinearity in a linear-nonlinear-Poisson (LNP) encoding model \citep{park2011active}, the receptive field of a neuron \citep{park2012bayesian}, or the parameters of a Hidden Markov Model (HMM) \citep{jha2022bayesian}.

However, implementing BAL for biological settings can be challenging, especially for real-time applications. Its applicability to real experiments is limited by two main drawbacks. Firstly, it requires computing an update of the posterior distribution of parameters after each time step,
and using it to compute the expected information gain from future experiments. This involves solving an optimization problem over a possibly high-dimensional stimulus space: current methods are either too time consuming, or only applicable to specific models. Secondly, to reduce computational complexity, classical implementations of BAL usually only optimize for the immediate next stimulus input. This classical myopic approach disregards all future observations in the experiment, and is thus possibly sub-optimal \citep{ryan2016review,drovandi2018improving}.

Our main contribution is to provide a general framework for online
active learning, called Efficient Sampling-Based Bayesian Active Learning (ESB-BAL). We use particle filtering, which is a highly versatile filtering method \citep{crisan2018nested}, for posterior computation; and propose a parallel computing implementation \citep{besard2018juliagpu,besard2019prototyping} for efficient posterior update and information computation. 
Whereas previous implementations of active learning either relied on time consuming Monte Carlo (MC) methods \citep{huan2013simulation,foster2019variational} or were only applicable to special cases, such as linear models or GLM \citep{lewi2009sequential}, our proposed solution is fast enough to be used in real-time biological experiments and can be applied to any state-space model.

To illustrate our method, we apply it to the problem of inferring the parameters of a chemical synapse with Short-Term Depression (STD). 
Upon the arrival of a presynaptic action potential, vesicles from a pool of $N$ independent release sites will fuse with the presynaptic plasma membrane with a probability $p$, each of these release events giving rise to a quantal current $q$ \citep{del1954quantal,katz1969release}. 
In addition, synaptic transmission is also dynamic. 
Short-term depression occurs when the inter-stimulation interval (ISI) is shorter than the time needed for synaptic vesicle replenishment \citep{tsodyks1998neural}. 
A synapse exhibiting STD can thus be described by its parameters $N$, $p$, $q$, and by its depression time constant. These parameters can be inferred using excitatory postsynaptic currents (EPSCs) recorded from the postsynaptic cell and elicited by stimulating the presynaptic axon. 
The accuracy of these estimates critically depends on the presynaptic stimulation times: if inter-stimulation intervals  are longer than the depression time constant, STD will not be precisely quantified. But if the stimulation frequency is too high, the pool of presynaptic vesicles will be depleted, leading to poor parameter estimates \citep{gontier2020identifiability,wieland2021structural}. Synaptic characterization is thus a relevant example application for ESB-BAL, as it requires careful tuning of the inputs $x_{1:T}$, but it is also a challenging one: computation needs to be faster than the typical ISI, which can be on the order of a few milliseconds. Using synthetic data, we show that our method allows to significantly reduce the uncertainty of the estimate in comparison to classically used non-adaptive stimulation protocols. 
We also show that the rate of information gain (in bit/s) of the whole experiment can be optimized by adding a penalty term for longer ISIs.
Lastly, we extend active learning to non-myopic designs. Using recordings from cerebellar mossy fiber to granule cell synapses from acute mouse brain slices, we show that our framework is sufficiently efficient for optimizing not only the immediate next stimulus, but rather the future stimuli in the experiment.

\section{Model}

\subsection{A general setting for Bayesian Active Learning}

\begin{figure}
\begin{fullwidth}

    \centering
    
     \begin{subfigure}[b]{0.37\linewidth}
         \centering
         \caption{}
         \includegraphics[width=\textwidth]{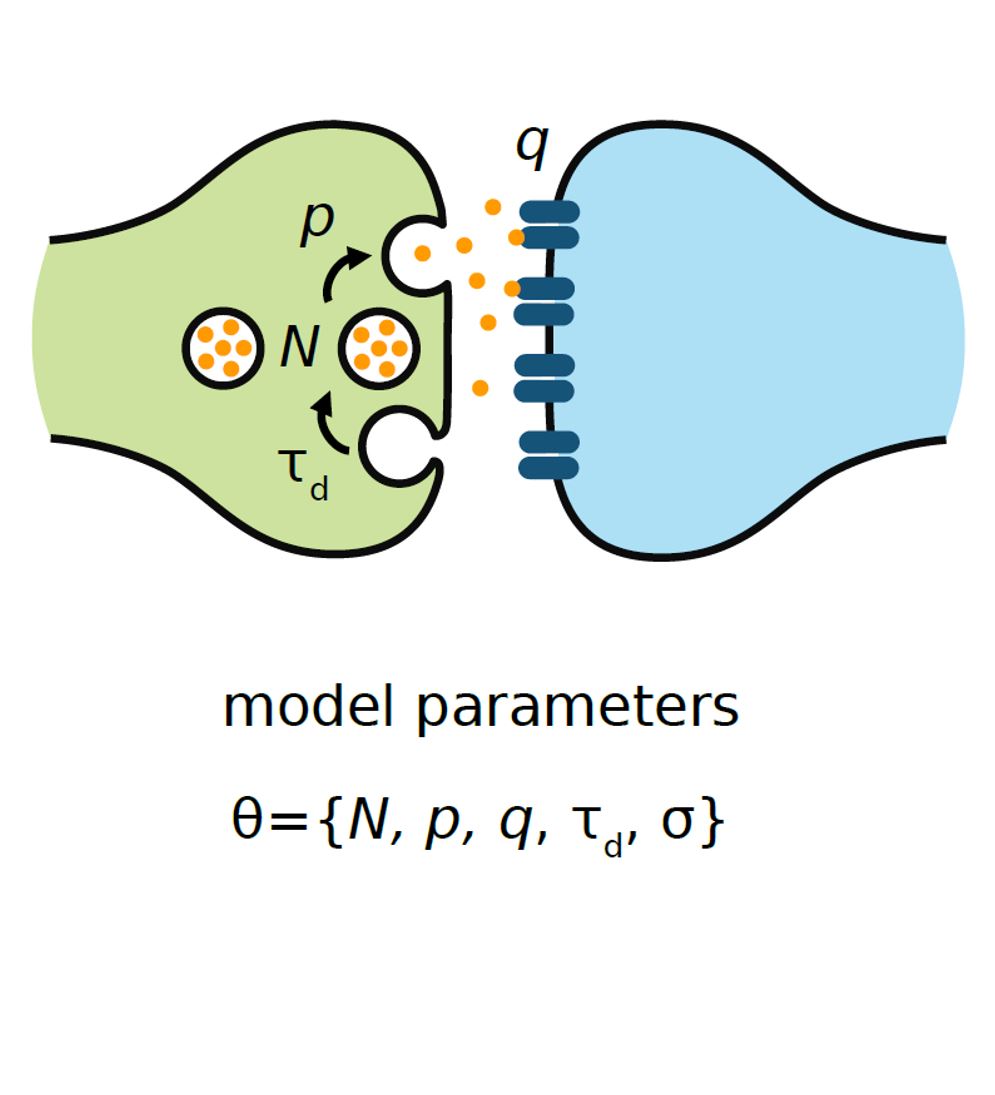}
         
     \end{subfigure}
     \begin{subfigure}[b]{0.55\linewidth}
         \centering
         \caption{}
         \includegraphics[width=\textwidth]{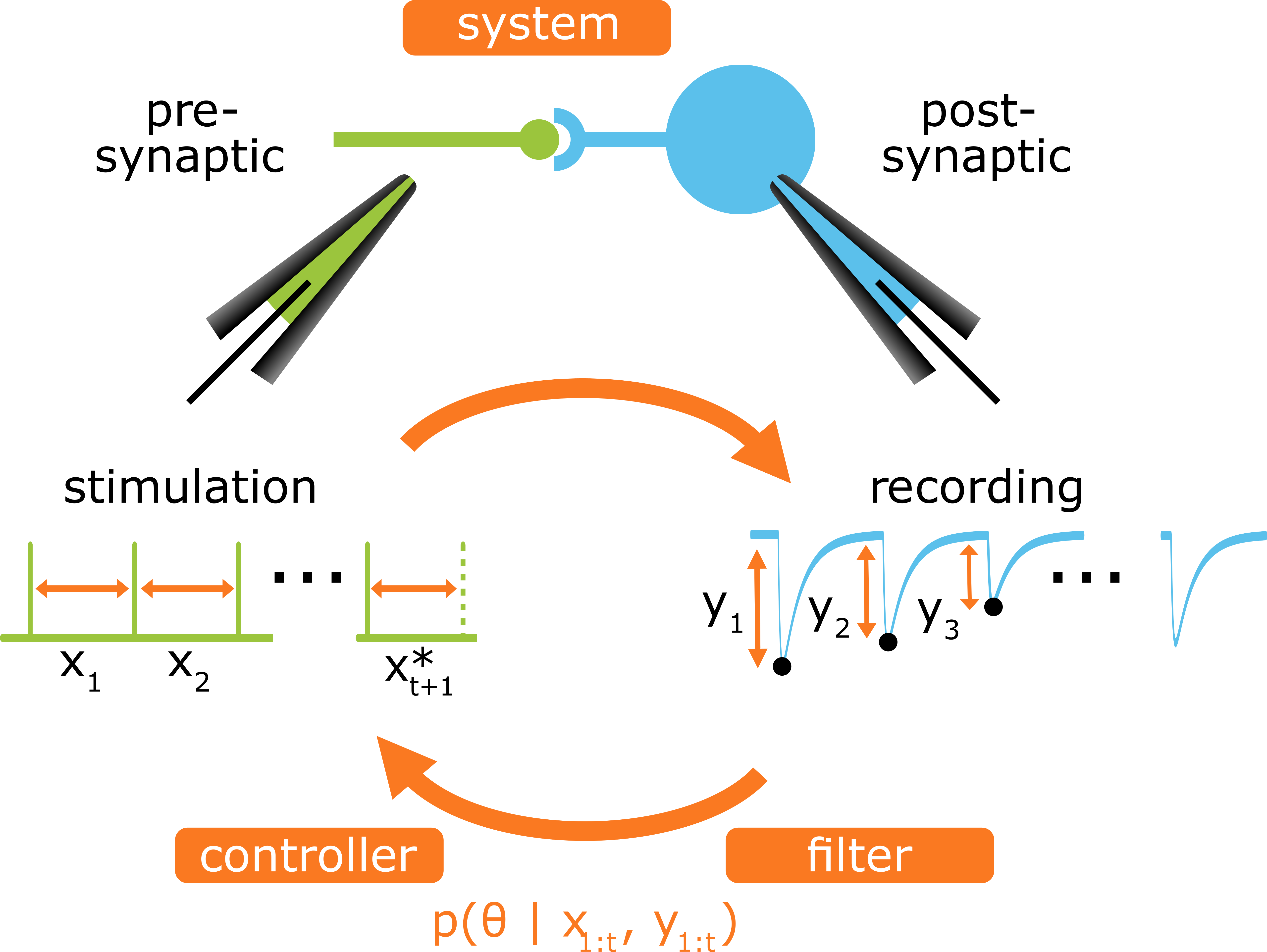}
         
     \end{subfigure}
 
    \caption{
  \textbf{(a)} Model of binomial synapse with STD. The presynaptic axon is stimulated to evoke an action potential. The input $x_t$ refers to the time interval since the previous stimulation, i.e., to the inter-spike interval $\Delta_t$. In chemical synapses, the presynaptic terminal is characterized by the presence of $N$ vesicles containing the neurotransmitter molecules, $n_t$ of them being in the readily-releasable state \citep{kaeser2017readily}. Upon the arrival of a presynaptic spike, these vesicles will stochastically fuse with the plasma membrane and release their neurotransmitters into the synaptic cleft. After spike $t$, $k_t$ vesicles (out of the $n_t$ available ones in the readily-releasable pool) release their neurotransmitters with a probability $p$. Neurotransmitters will bind to postsynaptic receptors: a single release event triggers a quantal response $q$. The total recorded postsynaptic current $y_t$ (i.e. the output of the system) is the sum of the effects of the $k_t$ release events. After releasing, vesicles are replenished with a certain time constant $\tau_D$, which determines short-term depression.
  \textbf{(b)} Bayesian Active Learning applied to biology.
  At each time step, the response of the \textbf{system} (e.g. here a synapse) to artificial stimulation is recorded. 
  This observation $y_t$ is used by the \textbf{filter} to compute the posterior distribution of parameters $p(\theta|x_{1:t},y_{1:t})$. 
  Given this posterior, the \textbf{controller} then computes the next input $x_{t+1}^*$ to maximize the expected gain of information of the next observation. In classical experiment design, the inputs $x_{1:T}$ are defined and fixed prior to the recordings.   
  }
    \label{fig:setup_synapse}
    \end{fullwidth}

\end{figure}

When using active learning in sequential experiments, three key elements need to be defined (Figure \ref{fig:setup_synapse} (b)):

\begin{enumerate}
    \item The \textbf{system} to be studied: it is described by a generative model $\mathcal{M}$, which parameters $\theta$ can be inferred from its observed responses $y_{1:T}$ to a set of $T$ input stimuli $x_{1:T}$. Given the stochastic nature of most systems studied in biology, the random variable $Y_{1:T}$ corresponding to the observations can take various values $y_{1:T}$ according to a distribution $p(y_{1:T}|x_{1:T},\theta)$. In our application example of BAL, the system will be a model of binomial neurotransmitter release (see Section \nameref{sec:system2}).
    \item A \textbf{filter} that computes the posterior distribution of the parameters given the previous inputs and observations $p(\theta| x_{1:t}, y_{1:t})$: after each new input $x_{t+1}$ and observation $y_{t+1}$, it is updated to obtain $p(\theta| x_{1:t+1}, y_{1:t+1})$ (see Section \nameref{sec:filter}).
    \item A \textbf{controller} that computes the next optimal input stimuli $x_{t+1}^*$ so as to maximize a certain utility function, which is often defined as the mutual information between the parameter random variable $\Theta$ and the response random variable $Y_{t+1}$ given the experimental inputs $I_{x_{t+1}}(\Theta;Y_{t+1}|h_t)$, where $h_t = (x_{1:t},y_{1:t})$ is the experiment history (see Section \nameref{sec:controller}).
\end{enumerate}

In synaptic characterization, inputs correspond to a set of $T$ stimulation times $x_{1:T}$ and observations correspond to recorded excitatory postsynaptic currents (EPSCs) $y_{1:T}$.
In case of successive experiments \citep{park2012bayesian}, the mutual information between the parameters and the next observation $Y_{t+1}$ conditioned on the experiment history $h_t$ is:

\begin{equation} \label{eq:mutual_info}
I_{x_{t+1}}(\Theta;Y_{t+1}|h_t) = H(\Theta|h_t) - H_{x_{t+1}} (\Theta|h_t,Y_{t+1}) 
\end{equation} 

where $H(\Theta|h_t)$ is the entropy of  $\Theta$ given the experiment history up to time step $t$:

\begin{equation}
H(\Theta|h_t) = \int d\theta \log p(\theta|h_t) p(\theta|h_t)
\end{equation} 

and 

\begin{equation} \label{eq:conditional_entropy}
H_{x_{t+1}} (\Theta|h_t,Y_{t+1})  = \int dy_{t+1} p(y_{t+1}|h_t,x_{t+1}) H_{x_{t+1}} (\Theta|h_t,Y_{t+1}=y_{t+1})
\end{equation}  

is the conditional entropy of  $\Theta$ given the future observation random variable $Y_{t+1}$.
Since the actual value of the future observation  is unknown, we take the average over $y_{t+1}$ of the conditional entropy $H_{x_{t+1}} (\Theta|h_,Y_{t+1}=y_{t+1})$ conditioned on a certain value $y_{t+1}$. As the predictive distribution  depends on the unknown parameters, we also have to take an average over $\theta$, using the current posterior distribution $p(\theta|h_t)$ at time $t$ \citep{lewi2009sequential}: 

\begin{equation}
p(y_{t+1}|h_t,x_{t+1}) = \int d\theta p(y_{t+1}|h_t,x_{t+1}, \theta)p(\theta|h_t)
\end{equation}

The goal of Bayesian active learning is to select the next stimulation to maximize the mutual information between the parameters and all future observations:

\begin{equation} \label{eq:next_time_non_myopic}
x_{t+1}^* = \argmax_{x_{t+1} \in \mathcal{S}_{t+1}} \max_{n} \max_{x_{t+2:t+n} \in \mathcal{S}_{t+2:t+n}} I_{x_{t+1:t+n}}(\Theta;Y_{t+1:t+n}|h_t)
\end{equation} 

 where $\mathcal{S}_{t+1}$ is the set of possible inputs at time step $t+1$ and $\mathcal{S}_{t+2:t+n}$ is the set of possible protocols for the stimulations from time step $t+2$ to $t+n$. This set of protocols includes all the stimulation constraints, e.g.  the remaining time of the experiment or the minimal inter-stimulation time.
Optimizing all future inputs is an intractable problem (especially for online applications), since the algorithmic complexity scales exponentially with the number of observations $n$. For this reason, BAL only optimizes for the next stimulus (an approach referred to as a \textit{myopic} design) (see Figure \ref{fig:setup_synapse} (b)):

\begin{equation} \label{eq:next_time}
x_{t+1}^* = \argmax_{x_{t+1} \in \mathcal{S}_{t+1}} I_{x_{t+1}}(\Theta;Y_{t+1}|h_t)
\end{equation} 

Different methods have been proposed to compute Eq. \ref{eq:next_time}. Monte Carlo (MC) methods \citep{huan2013simulation} or a variational approach \citep{foster2019variational} can be employed, but they usually require long computation times that can be impractical if the time between successive experiments is short. 
Closed-form solutions can be computed only for some special cases, such as linear models or GLM \citep{lewi2009sequential}.

\subsection{The system: a binomial model of neurotransmitter release}\label{sec:system2}

To illustrate our ESB-BAL framework, we apply it to the problem of estimating the parameters of a chemical synapse, represented as a state-space model with unobservable hidden states and input-dependent state transitions. A classical used model to describe the release of neurotransmitters at chemical synapses is called the binomial model \citep{katz1969release,barri2016quantifying,bird2016bayesian,gontier2020identifiability,stricker2003quantal,scheuss2001estimating,tsodyks1998neural}. 
According to this model, a synapse is described as an Input-Output Hidden Markov Model (IO-HMM) with the following parameters (units are given in square brackets, see also Figure \ref{fig:setup_synapse} (a)):

\begin{itemize}
    \item $N$ (the number of presynaptic independent release sites [-]);
    \item $p$ (their release probability upon the arrival of a presynaptic spike [-]);

    \item $\sigma$ (the standard deviation of the recording noise [A]);

    \item $q$ (the quantum of current elicited in the postsynaptic cell by one release event [A]);

    \item $\tau_D$ (the time constant of synaptic vesicle replenishment [s]).

\end{itemize}

The variables $n_t$ and $k_t$ represent, respectively, the number of available vesicles in the readily-releasable state at the moment of spike $t$ (with $0 \leq n_t \leq N$), and the number of vesicles (among $n_t$) released after spike $t$ (with $0 \leq k_t \leq n_t$).
For simplicity, we use the notations $p_{\theta}(\cdot) = p(\cdot|\theta)$ with $\theta = [N,p,q,\sigma,\tau_D]$, and $z_t := (n_t,k_t)$ to refer to the hidden variables at time step $t$. 

The probability of recording a set of $T$ EPSCs  $p_{\theta}(y_{1:T})$ is computed as the marginal of the joint distribution of the observations $y_{1:T}$ and the hidden variables $z_{1:T}$, i.e. $p_{\theta}(y_{1:T}) = \sum_{z_{1:T}} p_{\theta}(y_{1:T},z_{1:T})$, where the joint distribution $p_{\theta}(y_{1:T},z_{1:T}) = p_{\theta}(y_{1:T},n_{1:T},z_{1:T})$ is given by

\begin{equation}\label{eq:p_e}
p_{\theta}(y_{1:T},n_{1:T},k_{1:T}) = p_{\theta}(y_1|k_1)p_{\theta}(k_1|n_1)p_{\theta}(n_1) \prod_{t = 2}^{T} p_{\theta}(y_t|k_t)p_{\theta}(k_t|n_t)p_{\theta}(n_t|n_{t-1},k_{t-1},x_t)  
\end{equation}
where 

\begin{equation}
\label{eq:observation}
p_{\theta}(y_t|k_t) = \mathcal{N}(y_t;qk_t,\sigma^2) 
\end{equation}

is the emission probability, i.e. the probability to record output $y_t$ knowing that $k_t$ vesicles released neurotransmitter;
$p_{\theta}(k_t|n_t)$ is the binomial distribution and represents the probability that, given $n_t$ available vesicles, $k_t$ of them will indeed release neurotransmitter:

\begin{equation}
\label{eq:transition1}
p_{\theta}(k_t|n_t) =  \left( {\begin{array}{*{20}c} n_t \\ k_t \\ \end{array}} \right)p^{k_t} (1-p)^{n_t - k_t}
\end{equation}

Finally, $p_{\theta}(n_t|n_{t-1},k_{t-1},x_t)$ represents the process of vesicle replenishment. 
During the time interval $x_t$, each empty vesicle can refill with a probability $ \pi(x_t) = 1-\exp \left (-\frac{x_t}{\tau_D} \right )$ such that the transition probability $p_{\theta}(n_t|n_{t-1},k_{t-1},x_t)$ is given by:

\begin{equation}
\label{eq:refill_prob}
p_{\theta}(n_t|n_{t-1},k_{t-1},x_t) = \left( {\begin{array}{*{20}c} N-n_{t-1}+k_{t-1} \\ n_t-n_{t-1}+k_{t-1} \\ \end{array}} \right) \pi(x_t)^{n_t-n_{t-1}+k_{t-1}} (1-  \pi(x_t))^{N-n_t}
\end{equation}

One can note that $n_t = n_{t-1} - k_{t-1} + v_t$, where $v_t \sim \mathrm{Bin}({N-n_{t-1}+k_{t-1}, \pi(x_t)})$ is the number of refilled vesicles during the time interval $x_t$.
Eqs. \ref{eq:p_e} to \ref{eq:refill_prob} define the observation model of the studied system (see Figure \ref{fig:setup_synapse}), i.e. the probability of a set of observations $y_{1:T}$ given a vector of stimuli $x_{1:T}$ and a vector of parameters $\theta$.

\subsection{The filter: online computation of the posterior distributions of parameters}\label{sec:filter}

To be applicable for online experiments, the filtering block, which will compute the posterior distribution of parameters $p(\theta|h_t)$, needs to satisfy two requirements: 

\begin{enumerate}
    \item It must be sufficiently versatile to be applied to different systems and models; 
    \item It must be online (i.e. its algorithmic complexity should not increase with the number of observations) \citep{bykowska2019model}.
\end{enumerate}

A promising solution is particle filtering \citep{kutschireiter2020hitchhiker}, and especially the Nested Particle Filter (NPF) \citep{crisan2018nested}. 
This algorithm is asymptotically exact and purely recursive, thus allowing to directly estimate the parameters of a HMM as recordings are acquired. 

The NPF relies on two nested layers of particles to approximate the posterior distributions of both the static parameters $\theta$ of the model and of its hidden states $z_t$. 
A first outer filter with $M_{\rm out}$ particles is used to compute the posterior distribution of parameters $p(\theta|h_t)$, and for each of these particles, an inner filter with $M_{\rm in}$ particles is used to estimate the corresponding hidden states $z_t$ (so that the total number of particles in the system is $M_{\rm out} \times M_{\rm in}$).
After each new observation, these particles are resampled based on their respective likelihoods, hence updating their posterior distributions (Figure \ref{fig:posterior}).

The NPF was originally proposed for static HMMs, in which the state transition probability $p(z_{t+1}|z_t,\theta)$ is supposed to be constant. Here, we extend it to the more general class of Input-Output Hidden Markov Models (IO-HMMs, also called GLM-HMMs in neuroscience, see \cite{jha2022bayesian}), in which the state transition probability at time step $t$ depends on an external input $x_t$. For instance, state transition in our model of synapse is not stationary, but depends on the ISI $x_t$ (see Section \nameref{sec:binomial_release}).

The filter (Algorithm \ref{algo:1}) relies on the following approximation to recursively compute the likelihood of each particle. 
Once the observation $y_{t}$ has been recorded, the likelihood of particle $\theta_t^{i}$, with $i \in \{1,...,M_{\rm out}\}$, depends on

\begin{equation}
\label{eq:aaaa}
p(\theta_t^{i}|y_{1:t}) \propto p(y_t|y_{1:t-1},\theta_t^{i},x_t)p(\theta_t^{i}|y_{1:t-1})
\end{equation}

with

\begin{equation}
\label{eq:explanation_crisan}
p(y_t|y_{1:t-1},\theta_t^{i},x_t) = \sum_{z_{t-1:t}}  p(y_t|z_t,\theta_t^{i})p(z_t|z_{t-1},\theta_t^{i},x_t)
\\\
p(z_{t-1}|y_{1:t-1},\theta_t^{i})
\end{equation}

If the variance of the jittering kernel $\kappa$ (which mutates the samples to avoid particles degeneracy and local solutions, see \nameref{sec:methods_and_material}) is sufficiently small, and hence if $\theta_t^{i} \approx \theta_{t-1}^{i}$, the approximation $p(z_{t-1}|y_{1:t-1},\theta_t^{i}) \approx p(z_{t-1}|y_{1:t-1},\theta_{t-1}^{i})$ allows to recursively compute Eq. \ref{eq:aaaa}. In practice, the different terms in Eq. \ref{eq:explanation_crisan} are computed as such: $p(y_t|z_t,\theta_t^{i})$ corresponds to the \textit{Likelihood} step of Algorithm \ref{algo:1}; $p(z_t|z_{t-1},\theta_t^{i},x_t)$ corresponds to the \textit{Propagation} step; and $p(z_{t-1}|y_{1:t-1},\theta_t^{i})$ corresponds to the distribution of hidden states at time $t-1$.

Contrary to previous methods for fast posterior computation that were only applicable to specific models \citep{lewi2009sequential}, our filter can be applied to any state-space dynamical system, including non-stationary and input-dependent ones. 
Moreover, it does not require to approximate the posterior as a Gaussian nor require a time consuming (and possibly unstable) numerical optimization step, while being highly parallelizable and efficient \citep{besard2018juliagpu,besard2019prototyping}.

\subsection{The controller: computation of the optimal next stimulation time}\label{sec:controller}

The objective of experiment design optimization is to minimize the uncertainty of the estimates (classically quantified using the entropy) while reducing the cost of experimentation (defined as the number of required trials, samples, or observations). The optimal next stimulus $x_{t+1}^*$ that will maximize the mutual information (i.e. minimize the uncertainty about $\theta$ as measured by the entropy) can be written from Eqs. \ref{eq:mutual_info}, \ref{eq:conditional_entropy}, and \ref{eq:next_time} as
\begin{equation}\label{eq:no_simplification}
x_{t+1}^* = \argmin_{x_{t+1}\in \mathcal{S}_{t+1}} \int d\theta p(\theta|h_t) \int dy_{t+1}  p(y_{t+1}|h_t,x_{t+1},\theta) H_{x_{t+1}} (\Theta|h_t,Y_{t+1}=y_{t+1})
\end{equation}
Eq. \ref{eq:no_simplification} requires to compute two (possibly high-dimensional) integrals over $\theta$ and $y_{t+1}$, for which closed-form expressions only exist for specific models. 
To avoid long MC simulations, we propose to use mean-field computations and to replace integrals by point-based approximations. 
Firstly, instead of computing the full expectation over $p(\theta|h_t)$, we set $\theta$ to the mean posterior value $\hat{\theta}_t = \int d \theta p(\theta|h_t) \theta $, which can be conveniently approximated as $\hat{\theta}_t \approx \frac{1}{M_{\rm out}} \sum_{i=1}^{M_{\rm out}} \theta_t^i $. Eq. \ref{eq:no_simplification} thus becomes

\begin{equation}\label{eq:simplification1}
x_{t+1}^* \approx \argmin_{x_{t+1} \in \mathcal{S}_{t+1}}  \int dy_{t+1} \, p(y_{t+1}|h_t,x_{t+1},\hat{\theta}_t) H_{x_{t+1}} (\Theta|h_t,Y_{t+1}=y_{t+1})
\end{equation}

Depending on the nature of the studied system and on the time constraints of the experiment, different estimators can also be used, such as e.g. $\hat{\theta}_t =    \argmax_{\theta} p(\theta|h_t)$. Secondly, instead of computing the full expectation over the future observation, we set $y_{t+1}$ to its expected value; Eq. \ref{eq:no_simplification} thus becomes

\begin{equation}\label{eq:simplification2}
x_{t+1}^* \approx \argmin_{x_{t+1}  \in \mathcal{S}_{t+1}}  H_{x_{t+1}} (\Theta|h_t, Y_{t+1}=\mathbb{E}(Y_{t+1}|h_t,x_{t+1},\hat{\theta}_t))
\end{equation}

In the general case, $\mathbb{E}(Y_{t+1}|h_t,x_{t+1},\hat{\theta}_t)$ can be  computed using Bayesian Quadrature \citep{acerbi2018variational}. More specifically, for our model of a chemical synapse, an analytical formulation for the expected value $\mathbb{E}(Y_{t+1}|x_{1:t+1},\hat{\theta}_t)$ can be efficiently derived using mean-field approximations (see Section \nameref{sec:mean_field}).
For each candidate $x_{t+1}$ in a given finite set $\mathcal{S}_{t+1}$, the entropy $H (\Theta|h_t,x_{t+1}, Y_{t+1}=\mathbb{E}(Y_{t+1}|h_t,x_{t+1},\hat{\theta}_t))$ can be computed using Algorithm \ref{algo:1}. 

Finally, by assuming that the posterior distribution of $\Theta$ is well approximated by a Gaussian distribution (which is the case when there are sufficient observations \citep{paninski2005asymptotic}), its entropy can be estimated as $\frac{1}{2} \log |2 \pi e \Sigma_t|$, where $\Sigma_t$ is the covariance matrix of the particles $\{\theta_{t}^{i}\}_{1 \leq i \leq M_{\rm out}}$ \citep{jha2022bayesian}.

\section{Results}\label{sec:binomial_release}

\subsection{First setting: reducing the uncertainty of estimates for a given number of observations}\label{sec:first_setting}

From the experimentalist point of view, a highly relevant question is how to optimize the stimulation protocol such that the measured EPSCs are most informative about synaptic parameters.
Previous studies showed that some stimulation protocols are more informative than others, but ignored the temporal correlations of the number of readily-releasable vesicles \citep{costa2013probabilistic} or did not compute which protocol would be most informative \citep{barri2016quantifying}. In classical deterministic experiment protocols, the stimulation times $x_{1:T}$ are defined and fixed prior to the recordings. By contrast, active learning optimises the protocol on the fly as data are recorded.

\begin{figure}
\begin{fullwidth}
\begin{minipage}[c][][c]{.49\linewidth}
  \vspace*{\fill}
  \centering
  \caption*{\textbf{(a)}}
  \includegraphics[width=1\textwidth]{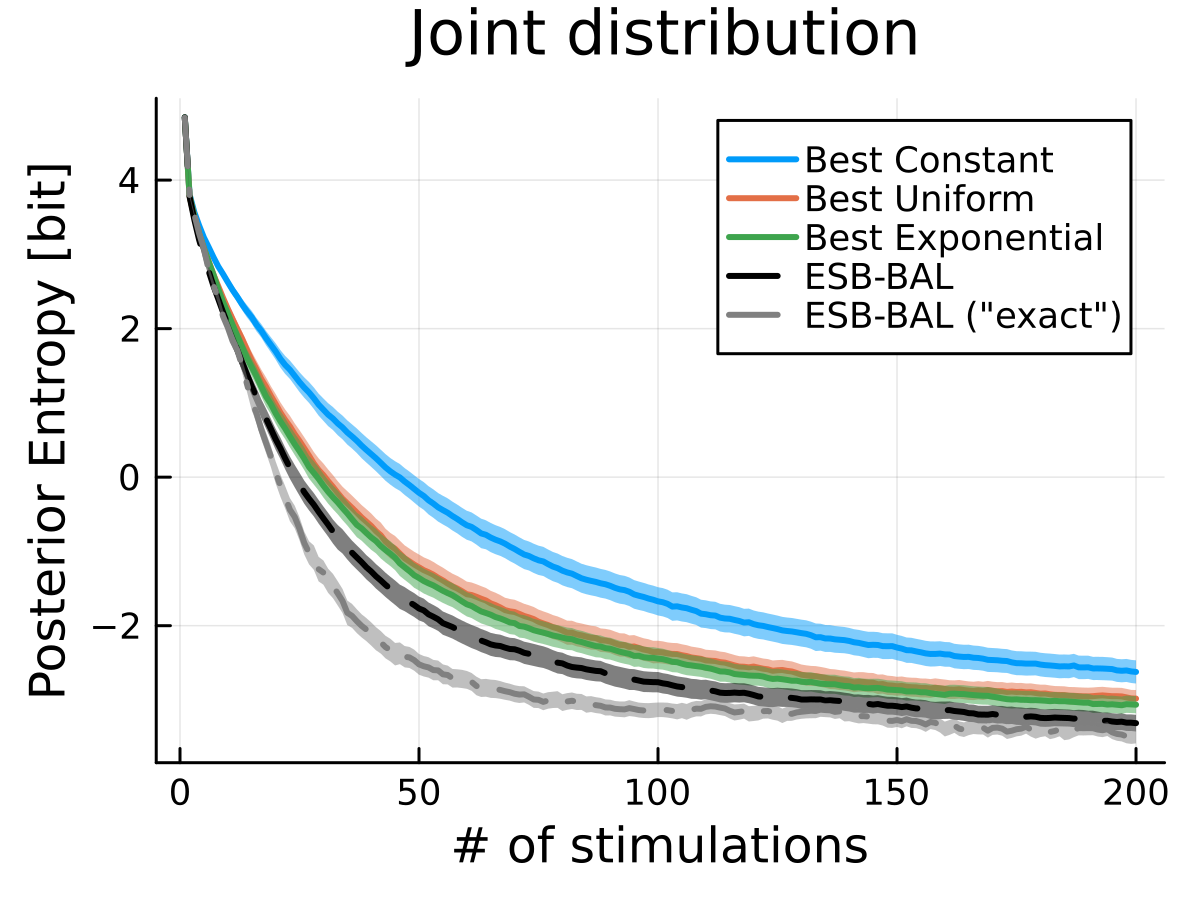}
\end{minipage}%
\begin{minipage}[c][][c]{.49\linewidth}
  \vspace*{\fill}
  \centering
  \caption*{\textbf{(b)}}
  \includegraphics[width=1\textwidth]{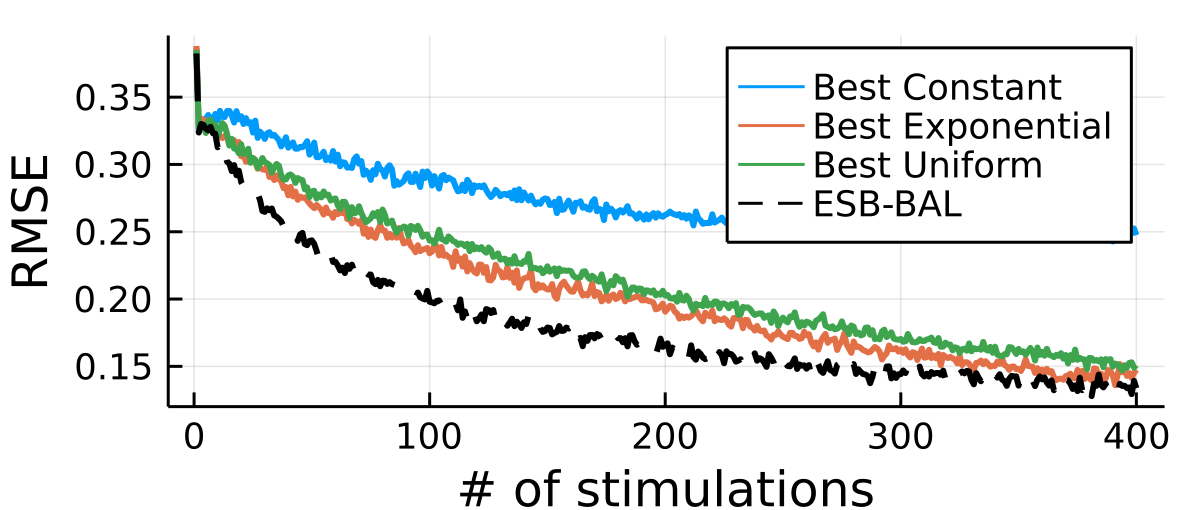}
  \par\vfill
  \caption*{\textbf{(c)}}
  \includegraphics[width=1\textwidth]{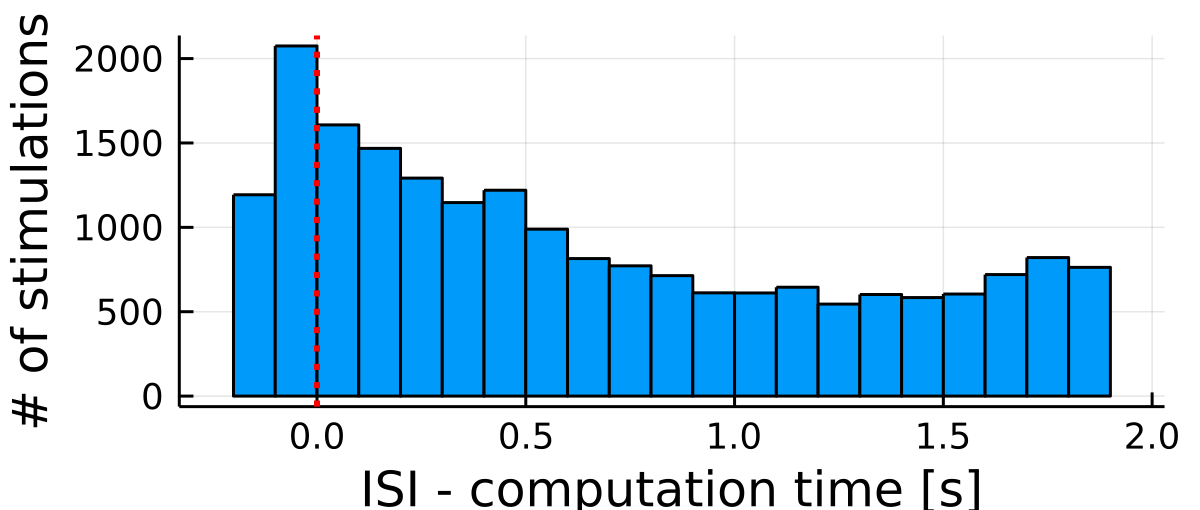}
\end{minipage}
\caption{\textbf{First setting: reducing the uncertainty of estimates for a given number of observations.} 
\textbf{(a)} Entropy of the posterior distribution of $\theta$ vs. number of observations for different stimulation protocols. Synthetic data were generated from a model of synapse with ground truth parameters $N^* = 7$, $p^* = 0.6$, $q^* = 1$ A, $\sigma^* = 0.2$ A, and $\tau_D^* = 0.25$s \citep{bird2016bayesian}. Gray dashed line corresponds to "exact" active learning, in the sense that $x_{t+1}^*$ is computed from Eq. \ref{eq:no_simplification} using MC samples instead of using Eq. \ref{eq:simplification2}. Traces show average over 100 independent repetitions. Shaded area: standard error of the mean. \textbf{(b)} RMSE for the same simulations. \textbf{(c)} Histogram of the differences between the ISI and the corresponding computation time for the ESB-BAL simulations.}
\label{fig:fig3}
\end{fullwidth}
\end{figure}

Results for a simulated experiment with ground-truth parameters $N^* = 7$, $p^* = 0.6$, $q^* = 1$ pA, $\sigma^* = 0.2$ pA, and $\tau_D^* = 0.25$s (i.e. the same set of parameters $\theta^*$ used in \cite{bird2016bayesian}) are displayed in Figure \ref{fig:fig3} (a).
Here, we compare ESB-BAL to three deterministic protocols:

\begin{itemize}
    \item in the \textit{Constant} protocol, the synapse is probed at a constant frequency, i.e. $ x_t = \rm cst$;
    \item in the \textit{Uniform} protocol, ISIs are uniformly drawn from a set $\mathcal{S}$ of candidates $x_t$ consisting of equidistantly separated values ranging from $x^{\rm min} = 0.005$s (i.e. one order of magnitude shorter than the shortest ISI used in \cite{barri2016quantifying}) to $x^{\rm max}$, i.e. $x_t \sim \mathrm{Uniform}([0.005,x^{\rm max}])$;
    \item finally, in the \textit{Exponential} protocol, ISIs are drawn from an exponential distribution with mean $\tau$. Such a protocol has been shown to provide better estimates of synaptic parameters compared to periodic spike trains with constant ISI \citep{barri2016quantifying,costa2013probabilistic}.
\end{itemize}

The efficiency of these deterministic protocols will depend on their respective parametrizations. To conservatively assess ESB-BAL, we optimize the values of $x_t$, $x^{\rm max}$, and $ \tau$ so that the \textit{Constant}, \textit{Uniform}, and \textit{Exponential} protocols have the best possible performance for the used ground-truth parameters $\theta^*$. Figure \ref{fig:results_optim2} shows the average final entropy decrease (i.e. the information gain) after 200 observations using
the \textit{Constant} (top), \textit{Uniform} (middle), or \textit{Exponential} (bottom) protocol, for different values of their
hyperparameters. These deterministic protocols (with their optimal respective parametrizations) are then compared to ESB-BAL.

 For the different protocols, the average (over 100 independent repetitions) joint differential entropy of the posterior distribution of parameters is plotted as a function of the number of observations (Figure \ref{fig:fig3} (a)).
ESB-BAL allows to  reduce the uncertainty (as measured by the entropy) of the parameter estimates for a given number of observations.
 It should be noted that it is compared to deterministic protocols whose respective hyperparameters have been optimized offline, knowing the value of $\theta^*$. In real physiology experiments, classical protocols
are non-adaptative and are defined using (possibly sub-optimal) default parameters. In contrast,
in active learning the protocol is optimized on the fly as data are recorded, and its performance
will not depend on a prior parametrization.
 ESB-BAL thus outperforms the best possible \textit{Constant}, \textit{Uniform}, and \textit{Exponential} protocols.
 
 We also verify that ESB-BAL does not lead to biased estimates of $\theta$, as its average RMSE outperforms that of other protocols (Figure \ref{fig:fig3} (b)); and that it is sufficiently fast for online applications, as computation time exceeds the ISI in only a small proportion of cases (Figure \ref{fig:fig3} (c)). Similar results can be observed for different sets of ground-truth parameters $\theta^*$ (Figure \ref{fig:fig8}) or when only optimizing for the entropy of a specific parameter (Figure \ref{fig:fig7}).
 
Finally, in Figure \ref{fig:fig3} (a), ESB-BAL (black dashed line) is compared to exact active learning (gray dashed line), in which Eq. \ref{eq:no_simplification} is computed exactly using MC samples. 
Samples to compute the expectation over $\theta$ are drawn from $p(\theta|h_t)$, whereas samples used to compute the expectation over $y_{t+1}$ are drawn from the generative distribution $p(y_{t+1}|h_t,x_{t+1},\theta)$ (Eqs. \ref{eq:observation}, \ref{eq:transition1}, and \ref{eq:refill_prob}).
This shows that the approximations used in Algorithm \ref{algo:2} to make active learning online have only a small effect on performance.

\subsection{Second setting: reducing the uncertainty of estimates for a given experiment time}

Active learning allows, for a given number of observations, to improve the reliability of the estimated parameters. 
However, in its classical implementation, only the next stimulus input is optimized, disregarding all future observations in the experiment. 
This myopic approach is thus sub-optimal. 
Moreover, neurophysiology experiments are not only constrained by the number of observations, but also by the total time of the experiment. 
Since cell viability and recording stability may become limiting during an experiment, the total time of an experimental protocol 
$\sum_{t=1}^T x_t$ also needs to be accounted for.
Here, to account for the total time of the experiment, and to globally optimize the information gain per unit of time, we propose to modify the classical formulation of active learning (Eq. \ref{eq:no_simplification}) by adding a penalty term for longer ISIs:

\begin{equation}\label{eq:penalty}
x_{t+1}^{* (\eta)} = \argmin_{x_{t+1} \in \mathcal{S}_{t+1}} \left \{ \eta x_{t+1} + \int d\theta p(\theta|h_t) \int dy_{t+1}  p(y_{t+1}|h_t,x_{t+1},\theta) H_{x_{t+1}} (\Theta|h_t,Y_{t+1}=y_{t+1}) \right \}
\end{equation}

\begin{figure}
\begin{fullwidth}

    \centering
    
     \begin{subfigure}[b]{0.45\linewidth}
         \centering
         \caption{}
         \includegraphics[width=\textwidth]{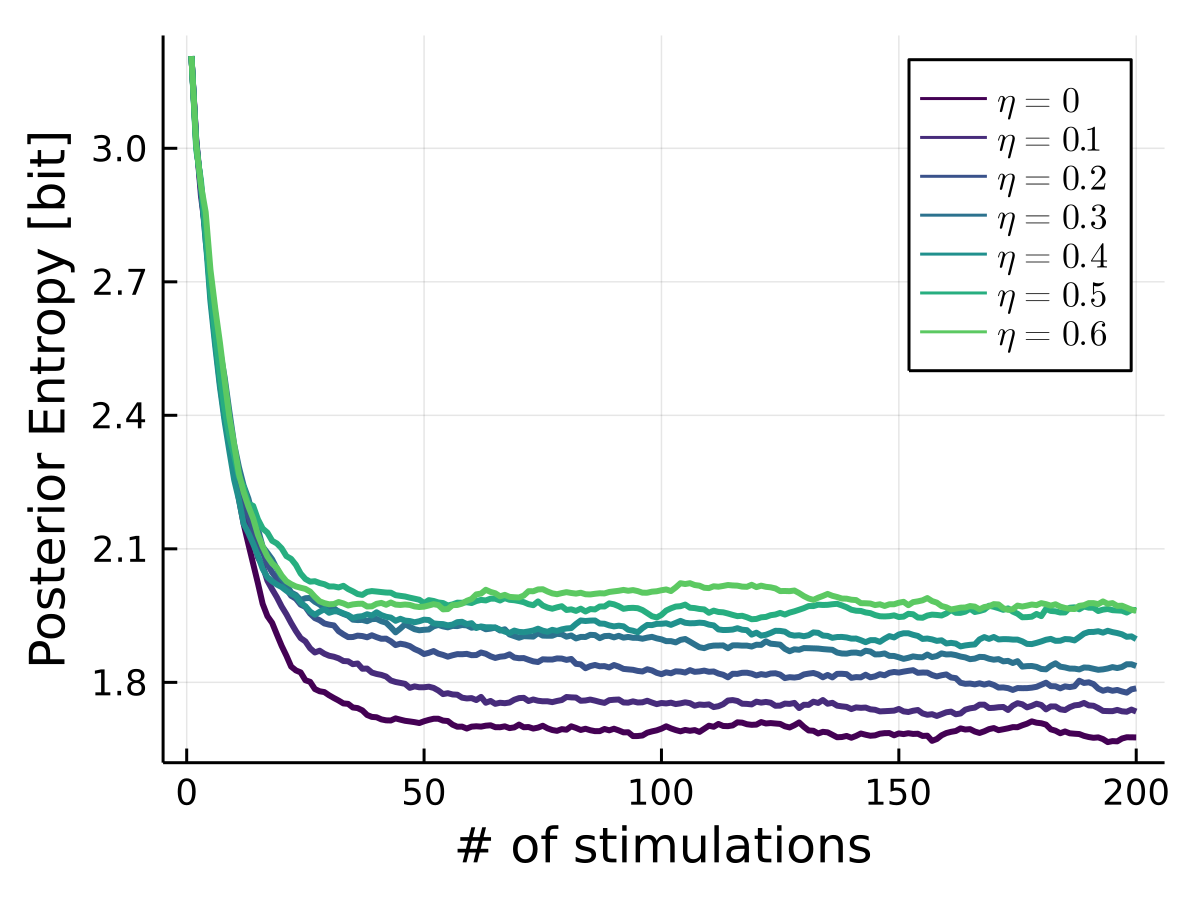}
         
     \end{subfigure}
     \begin{subfigure}[b]{0.45\linewidth}
         \centering
         \caption{}
         \includegraphics[width=\textwidth]{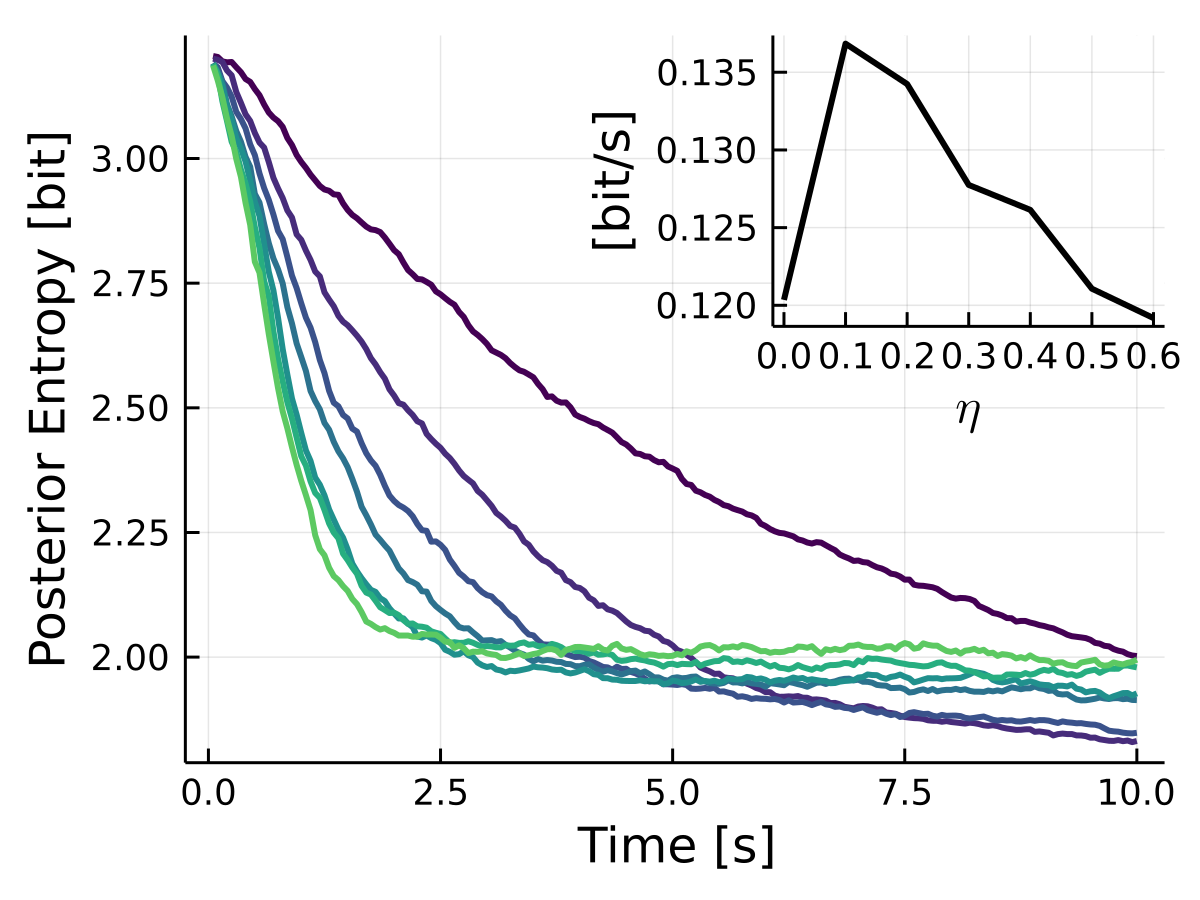}
         
     \end{subfigure}
 
    \caption{\textbf{Second setting: reducing the uncertainty of estimates for a given experiment time}
    (effect of penalizing long ISIs on parameter estimates uncertainty and rate of information gain). \textbf{(a)} Entropy of the posterior distribution of $\tau_D$ vs. number of observations for different values of $\eta$ in Eq. \ref{eq:penalty}. Same settings as in Figure \ref{fig:fig8}. \textbf{(b)} Same results, but displayed as a function of time. Inset: slope of the entropy vs. time curves (i.e. information rate) vs. $\eta$ after 10 seconds.}
    \label{fig:fig4}
    \end{fullwidth}

\end{figure}

The effect of the penalty weight $\eta$ on the entropy of the posterior distribution of $\tau_D$ is displayed in Figure \ref{fig:fig4} (a). As expected, adding a penalty term to Eq.
\ref{eq:no_simplification} reduces the precision of the inferred parameter. 
The loss of information gain increases with the penalty weight $\eta$. 
However, increasing $\eta$ also increases the speed of information gain, as seen in Figure \ref{fig:fig4} (b). 
Depending on the available time for the experiment, it is thus possible to tune $\eta$ so as to find a trade-off between long-term precision (Figure \ref{fig:fig4} (a)) and information rate (Figure \ref{fig:fig4} (b)).

\subsection{Third setting: batch optimization and application to neural recordings}

\begin{figure}
\begin{fullwidth}

     \begin{subfigure}[b]{0.95\linewidth}
         \centering
         \caption*{\textbf{(a)}}
  \includegraphics[width=1\textwidth]{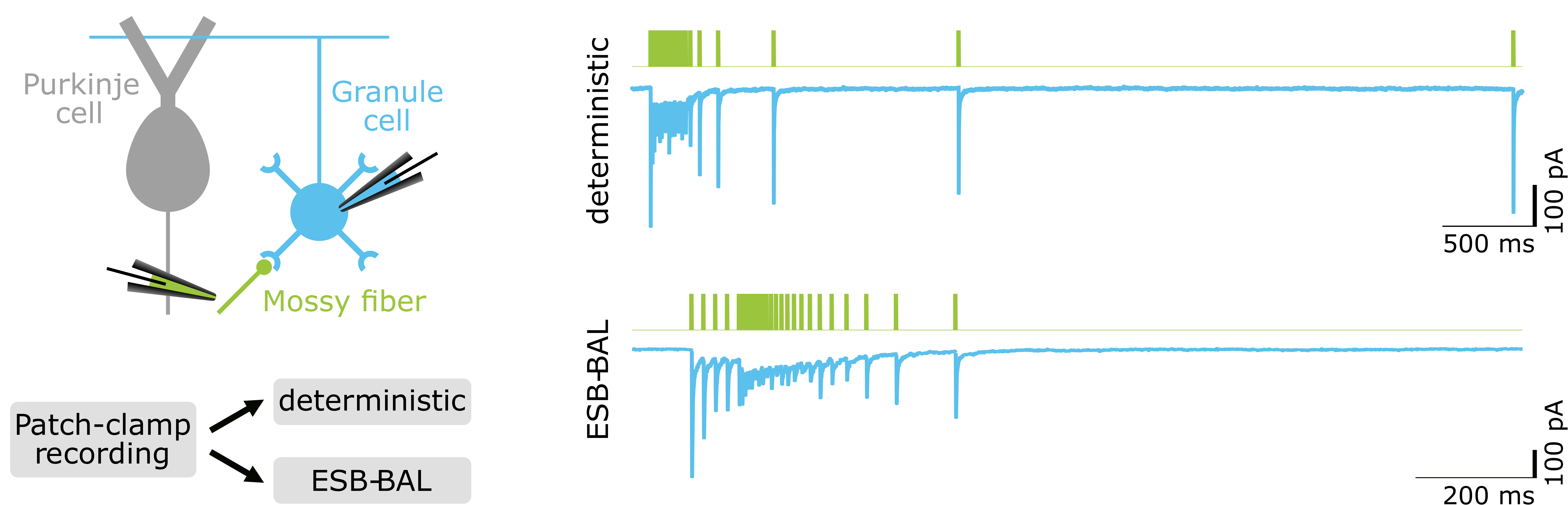}
  \end{subfigure}
  
  \begin{minipage}[c][][c]{.49\linewidth}
  \vspace*{\fill}
  \centering
  \caption*{\textbf{(b)}}
  \includegraphics[width=1\textwidth]{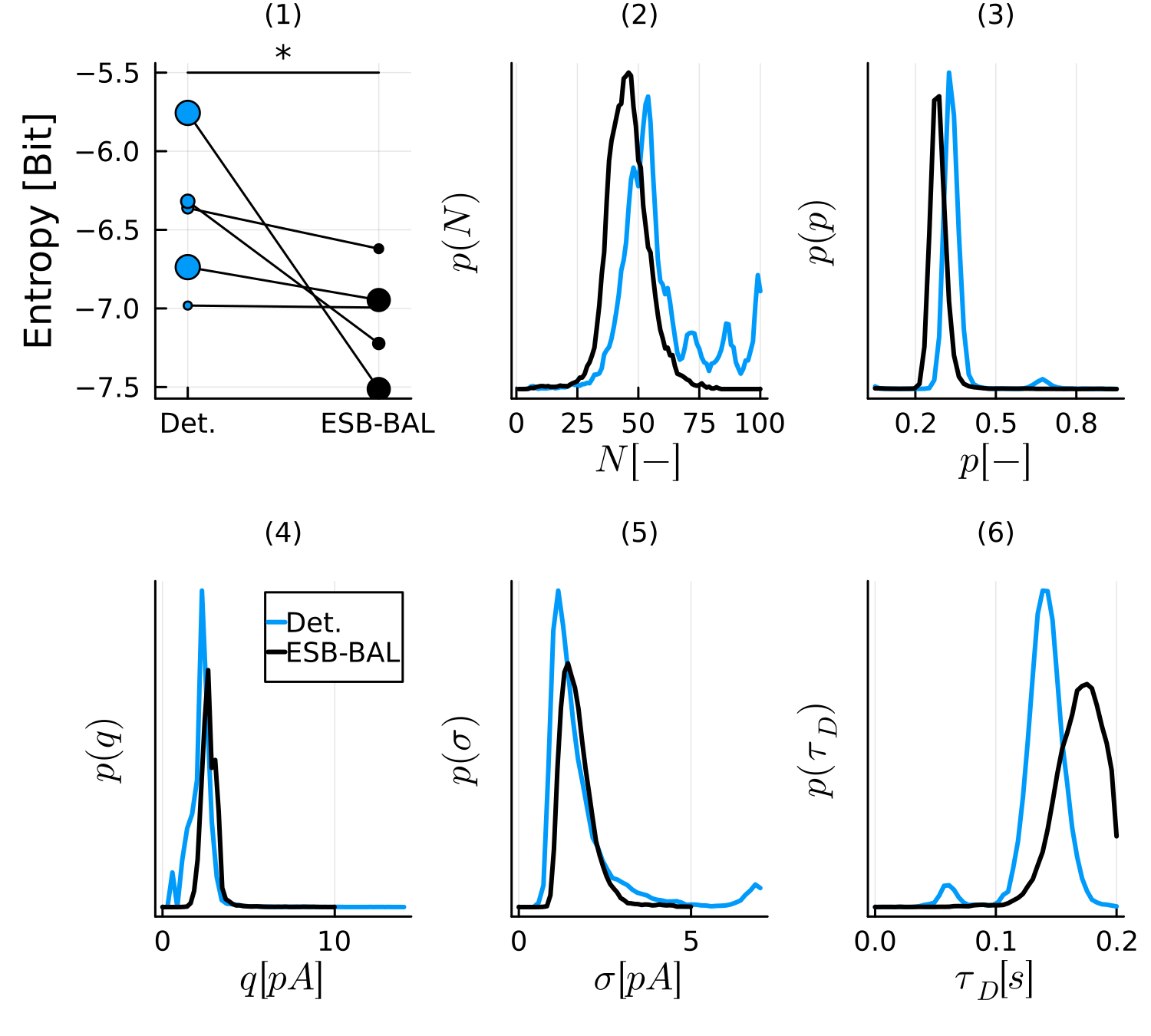}
\end{minipage}%
\bigskip{}
\bigskip{}
\begin{minipage}[c][][c]{.49\linewidth}
  \vspace*{\fill}
  \centering
  \caption*{\textbf{(c)}}
  \includegraphics[width=1\textwidth]{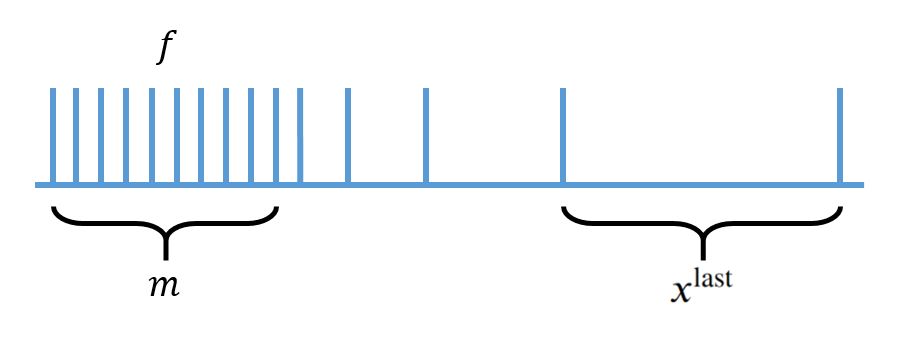}
  \par\vfill
  \caption*{\textbf{(d)}}
  \includegraphics[width=1\textwidth]{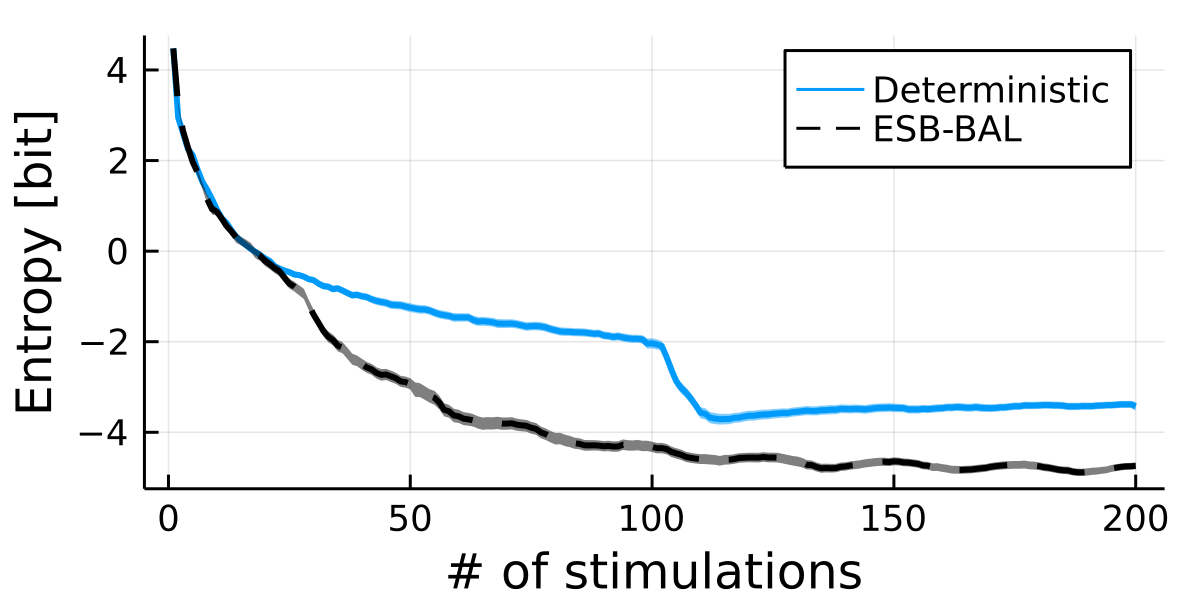}
\end{minipage}

     \caption{\textbf{Third setting: batch optimization and application to neural recordings}. 
     \textbf{(a)} Left: 5 mossy fiber to granule cell synaptic connections from acute cerebellar slices of mice were studied. Each of them was stimulated using both deterministic protocols and ESB-BAL. Right: examples of postsynaptic current traces recorded from a  granule cell upon extracellular mossy fiber stimulation.
     \textbf{(b)} Panel (1): Entropy at the end of simulation protocols for the 5 studied synapses. Markers' sizes are proportional to the number of observations obtained for each synapse. Test: weighted regression analysis ($p=0.01945$).
     Panels (2) to (6): Marginal posterior distributions for an example cell using either a deterministic protocol (blue) or ESB-BAL (black).
     \textbf{(c)} Schematic of how elements in $\mathcal{S}_{t+1:t+n}$ in Algorithm \ref{algo:3} are defined. They are chosen to span 3 parameters: the number $m < n$ of spikes in the tetanic stimulation phase, the frequency $f$ of spikes in the tetanic stimulation phase, and the duration of the final recovery ISI $x^{\rm last}$. 
     \textbf{(d)} Simulated experiment with ground-truth parameters $N^* = 47$, $p^* = 0.27$, $q^* = 2.65$ pA, $\sigma^* = 1.32$ pA, and $\tau_D^* = 0.17$s (i.e. the MAP values from the recordings shown in (b)).
     }
      \label{fig:fig5}
\end{fullwidth}

\end{figure}

To reduce computational complexity, classical implementations of sequential experiment design usually only optimize for the immediate next observation. However, it might be critical for some systems to optimize not only the next stimulus, but rather the next $n$ stimuli of the experiment altogether (see Eq. \ref{eq:next_time_non_myopic}) \citep{ryan2016review,drovandi2018improving}. Synaptic characterization is a telling example: indeed, STD can only be observed for specifically organized batches of stimulation times. When probing the presynaptic cell, neuroscientists usually use repetitions of a spike train (Figure \ref{fig:fig5} (c)) consisting of a tetanic stimulation phase (sustained high-frequency stimulation used to deplete the presynaptic vesicles) followed by recovery spikes at increasing ISIs to probe the STD time constant \citep{markram1998potential}. These spike trains (especially the duration and frequency of the tetanic phase, and the ISI between recovery spikes) are usually not optimized, and are held constant throughout an entire experiment.

Here, we show that ESB-BAL can be used to extend active learning to non-myopic designs, and to optimize the $n$ next input stimuli. Algorithm \ref{algo:3}, which is a generalization of Algorithm \ref{algo:2}, is used to select the next batch of $n$ stimuli $x_{t+1:t+n}^*$ in a set of candidate batches $\mathcal{S}_{t+1:t+n}$. Every $n$ observations, $H_{x_{t+1:t+n}} (\Theta|h_t,Y_{t+1:t+n})$ is computed using $n$ iterations of the filter (i.e. Algorithm \ref{algo:1}), in order to pick the optimal next batch $x_{t+1:t+n}^*$ that minimizes the quantity $H_{x_{t+1:t+n}} (\Theta|h_t,Y_{t+1:t+n})$ (i.e. the posterior entropy over the parameters at time step $t+n$ given all observations up to time $t$):

\begin{equation}
x_{t+1:t+n}^* = \argmin_{x_{t+1:t+n} \in \mathcal{S}_{t+1:t+n}}  H_{x_{t+1:t+n}} (\Theta|h_t,Y_{t+1:t+n})
\end{equation}

We validate our method by applying it to EPSC recordings from acute mouse cerebellar slices (Figure \ref{fig:fig5} (a)): 5 mossy fiber to granule cell synaptic connections were studied. Each of them was stimulated using both ESB-BAL and different deterministic protocols: in deterministic protocols, the presynaptic cell is stimulated using a repetitive train stimulation consisting of either 20 or 100 stimuli at either 100 Hz or 300 Hz followed by 6 recovery pulses at increasing intervals.

For each stimulation protocol, the posterior distribution of the parameters was computed offline using the Metropolis-Hastings algorithm. Figure \ref{fig:fig5} (b1) shows the entropy at the end of the different stimulation protocols for the 5 studied synaptic connections. Each synapse was stimulated using several protocols (once with ESB-BAL, and at least once with a deterministic one), each having possibly different number of observations $T_1,T_2,T_3 \dots$. Hence, for each synapse, only the first $T= {\rm min} \{T_1,T_2,T_3 \dots\}$ observations of each protocol were kept, to compare them for the same number of observations. In case a synapse was stimulated with several deterministic protocols, their respective entropies were averaged to only get one point in Figure \ref{fig:fig5} (b1).
Panels (2) to (6) show the marginal posterior distributions for parameters of an example synapse, obtained using either ESB-BAL or a deterministic protocol (consisting of repetitions of the same train of 20 stimuli at 100Hz followed by 6 recovery spikes). Our results show that batch optimization via ESB-BAL significantly outperforms deterministic stimulation.

Candidate batches of stimulation times in $\mathcal{S}_{t+1:t+n}$ are parametrized with a low-dimensional parametrization to span different durations and frequencies for the tetanic phase, and different ISIs between the recovery spikes (see Figure \ref{fig:fig5} (c) and \nameref{sec:methods_and_material}).
They are chosen to span 3 parameters: the number $m < n$ of spikes in the tetanic stimulation phase, the frequency $f$ of spikes in the tetanic stimulation phase, and the duration of the final recovery ISI $x^{\rm last}$.
The remaining $n-(m+1)$ spikes are then distributed geometrically between the end of the tetanic phase and the penultimate spike.

\section{Discussion}

\textbf{Summary}: 
We developed a method called Efficient Sampling-Based Bayesian Active Learning (ESB-BAL). Using particle filtering, ESB-BAL selects the next experimental design to maximize the mutual information between the output of the experiment and the constants of the studied system. To validate it, we apply ESB-BAL to the problem of estimating the constants of a chemical synapse from its postsynaptic currents evoked by presynaptic stimulations. After each new observation, the optimal next stimulation time can be computed using ESB-BAL. Using synthetic data and synaptic whole-cell patch-clamp recordings in cerebellar brain slices, we show that our method is efficient and fast enough to be used in real-time biological experiments and can  reduce the uncertainty of inferred parameters.

For illustrative purposes, we applied ESB-BAL to the specific problem of estimating the parameters characterizing a chemical synapse. However, we argue that our framework is sufficiently general and efficient to be applicable to a broad range of systems and domains of research. Especially, our extension of the Nested Particle Filter can be applied to any state-space system, even time-variant ones. Moreover, as the Nested Particle Filter is robust to time-varying parameters and model uncertainties \citep{crisan2018nested}, we believe that our proposed solution will be especially relevant for neurophysiology experiments or for clinical applications, for instance for optimizing Deep Brain Stimulation (DBS) for the treatment of Parkinson's Disease \citep{tinkhauser2021controlling,care2022impact}.

We expect active learning to be particularly beneficial to neurophysiology experiments involving live cells or subjects. By reducing the number of samples required to obtain a certain result, or by improving the efficiency of information gain, we can reduce the cost of the experiment and the need for animal subjects\footnote{A possible negative impact would be that improving the relative efficiency of neurophysiology experiments may lead to a larger field of applications and therefore a larger demand for animal experiments, analogously to Jevons Paradox \citep{jevons1862coal}.}.

\textbf{Limitations}: Our approach has some room for improvements. 
An evident drawback of using particle filtering is that it requires a very large number of particles to provide low variance estimates, as the approximation error only decreases with the square root of the number of particles. 
Moreover, future experimental work should focus on implementing ESB-BAL for different and more complicated models of a chemical synapse, including for instance short-term facilitation \citep{pfister2010synapses,costa2013probabilistic,barri2016quantifying,bird2016bayesian,gontier2020identifiability} or vesicle content variability \citep{bhumbra2013reliable,soares2019parsing}.
Finally, future theoretical work should focus on obtaining results on the convergence of the estimators when using active learning. 
When observations are independent and identically distributed (i.i.d.), active learning will give an unbiased estimate of the parameters, whose variance will decrease with the number of observations \citep{paninski2005asymptotic}.
Such theoretical results lack for systems with correlated outputs (such as the EPSCs in the studied synapse model), possibly leading to information saturation \citep{moreno2014information} or biased estimates.

For some recordings in Figure \ref{fig:fig5} (a), the benefit of using ESB-BAL instead of a deterministic protocol might seem non-significant. For some synaptic connections (e.g. synapse 3), ESB-BAL even yields a higher entropy for the posterior distribution of $\theta$. Different explanations can be put forward. Firstly, it is possible that the classically used deterministic protocols (20 stimuli at 100Hz followed by 6 recovery spikes at increasing ISIs, see Figure \ref{fig:fig5} (b)) are already well informative about the synaptic parameters. For these protocols, the tetanic stimulation phase and the long inter-sweep interval allow to estimate the value of the hidden states $n_t$ and $k_t$ with a high accuracy, which facilitates the estimation of the synaptic parameters \citep{barri2016quantifying}. Moreover, the recovery spikes at varying ISIs are known to be more informative about the synapse's dynamics than a constant stimulation frequency \citep{costa2013probabilistic}. Secondly, the small benefit of using ESB-BAL for some synaptic connections might be due to a model mismatch issue, as the model defined by Eqs. \ref{eq:p_e} to \ref{eq:refill_prob} might not exactly represent the ground-truth mechanisms of the studied synapses. Although mossy fiber to granule cell synapses are believed to be good examples of depressing synapses, our simplified model disregards several aspects of synaptic transmission such as facilitation, postsynaptic saturation, or presynaptic vesicles heterogeneity \citep{ritzau2014ultrafast,sargent2005rapid,saviane2006fast}. These assumptions might explain why ESB-BAL performs more consistently in simulations than in real experiments.

\textbf{Future work}: An interesting area of future research would be to formulate Optimal Experiment Design as an optimal control problem, using the framework of the Bellman equation \citep{bellman1966dynamic,sutton2018reinforcement}. 
This multi-stage optimization problem could be solved exactly by defining the associated Bellman equation, in which $I(\Theta;Y_{1:T})$ is the objective function, current observation $y_{t}$ is the state, input $x_{t}$ is the control, and where the optimal policy determines the next input $x_{t+1}$. This approach would allow to account for the remaining available experimental time.

Bayesian Active Learning is an efficient framework for solving the problem of optimal experiment design for parameters inference. Its goal is, for a given generative model $\mathcal{M}$,  to optimize the accuracy of the estimates of the parameters $\theta$ of $\mathcal{M}$, i.e. to minimize the entropy of the posterior distribution $p(\theta|x_{1:T},y_{1:T},\mathcal{M})$. But it is also possible to extend optimal experiment design to model selection: in this setting, the goal is to maximize the discriminability between competing candidate models, i.e. to minimize the entropy of $p(\mathcal{M}|x_{1:T},y_{1:T})$. Different schemes for OED for model selection have been proposed (see \cite{gontier2022statistical} for a discussion), but their computational complexity is a major impediment to their concrete applicability. An interesting future application of ESB-BAL would be to extend it to optimal model selection.

Overall, we expect our proposed solution to pave the way towards better estimates of stochastic models in neuroscience, more efficient training in machine learning, and more systematic and automated experimental designs.


\textbf{Conclusion}: When designing an experiment in physiology, or when training a model on data in machine learning, it is common to choose a priori a fixed set of inputs to the studied system.
The use of such non-adaptive, non-optimized protocols often leads to a large variance of the estimated parameters, even when using a large number of trials or data points. 
Bayesian active learning is an efficient method for optimizing these inputs, but exact solutions are often intractable and not applicable to online experiments. Here, we introduce ESB-BAL, a novel framework combining particle filtering, parallel computing, and mean-field theory. ESB-BAL is general and sufficiently efficient to be applied to a wide range of settings. 
We use it to infer the parameters of a model of synapse: for this specific example, computation time is a critical constraint, since the typical ISI is shorter than 1s, and because several future inputs need to be optimized together. Using synthetic data and neural recordings, we show that our method has the potential to significantly improve the precision and speed of model-based inferences.

\clearpage

\section{Methods and Materials}\label{sec:methods_and_material}

\subsection{Notations}

\begin{table}[ht!]
\caption{\label{tab:example} Notations.}
\begin{tabular}{l l}    \toprule
      \underline{\textbf{Indices:}}    \\

      $t \leq T$ & Number of observations\\
      $i \leq M_{\rm{out}}$ & Number of outer particles\\
      $j \leq M_{\rm{in}}$ & Number of inner particles\\

      \underline{\textbf{Parameters:}}  \\
      $N$ & Number of presynaptic independent release sites [-]\\
      $p$ & Release probability upon the arrival of a presynaptic spike [-]\\
      $q$ & Quantum of postsynaptic current elicited by one release event [A]\\

      $\sigma$ & Standard deviation of the recording noise [A]\\

      $\tau_D$ & Time constant of synaptic vesicle replenishment [s]\\
      \underline{\textbf{Random variables:}}  \\
      $\theta$ & Vector of unknown parameters \\
      $Y_t$ & Output of the system at time $t$ \\
      
      \underline{\textbf{Functions:}}  \\
      $p_{\theta}(\cdot)$ & Probability distribution conditioned on $\Theta = \theta$ \\
      
      $I(\cdot;\cdot)$ & Mutual information  \\
      $H(\cdot)$ & Differential entropy \\
      
      \underline{\textbf{Others:}}  \\
      $x_t$ & Input to the system at time step $t$ ($t^{th}$ inter-spike interval) \\
      $y_t$ & Recording at time step $t$ ($t^{th}$ EPSC amplitude) \\
      $h_t$ & History of observations ($x_{1:t},y_{1:t}$) \\ 
      $\mathcal{M}$ & Generative model of the studied system \\
      $n_t$ & Number of vesicles in the readily-releasable state immediately before spike $t$ \\ 
      $k_t$ & Number of released vesicles after spike $t$ \\
      \bottomrule
     \end{tabular}
    \end{table}

\clearpage

\subsection{Bayesian Active Learning}

For a fixed model $\mathcal{M}$, the goal of BAL is to optimize the accuracy of the estimates of its parameters $\theta$, i.e. to minimize the entropy of the posterior distribution $p(\theta|x,y)$ (see \cite{lindley1956measure,huan2013simulation} for a detailed discussion). The utility $\mathcal{U}(x,y)$ of a given experimental protocol $x$ and of a data set $y$ can be either defined as the gain in Shannon information between the prior and the posterior distribution of the parameters $\theta$, as suggested in \cite{lindley1956measure}:

\begin{equation} \label{first_def_U}
    \mathcal{U}(x,y) = \int d\theta \log p(\theta |x,y)p(\theta |x,y) - \int d\theta \log p(\theta)p(\theta)
\end{equation}

$\mathcal{U}(x,y)$ can also be defined as the Kullback-Leibler divergence between the prior and the posterior:
\begin{equation} \label{second_def_U}
    \mathcal{U}(x,y) = D_{KL}(p(\theta |x,y)||p(\theta))
\end{equation}

The expected utility $\mathcal{U}(x)$ of a protocol $x$ is finally the expected value of $\mathcal{U}(x,y)$ under $p(y|x)$, which yields the same result under \eqref{first_def_U} and \eqref{second_def_U}:

\begin{equation} \label{first_expression_U}
    \mathcal{U}(x) = \int dy \int d\theta  \log p(\theta |x,y) p(\theta, y|x) - \int d\theta \log p(\theta) p(\theta)
\end{equation}

It is worth noting that $\mathcal{U}(x)$ is actually the mutual information between $Y$ and $\Theta$. Indeed, Eq. \eqref{first_expression_U} can be rewritten as

\begin{equation}
\mathcal{U}(x) = \int d\theta \int dy \log p(\theta |x, y) p(\theta, y|x) + H(\theta) = H_x(Y)+H(\theta)-H_x(Y,\theta)
\end{equation}

which yields $\mathcal{U}(x) = I_x(\Theta;Y)$. Different MCMC-based methods to compute $\mathcal{U}(x)$ are described in \cite{huan2013simulation}. 

\subsection{Particle Filtering for synaptic characterization}\label{sec:appendix_Particle_Filtering_for_synaptic_characterization}

 \textbf{Initialisation}: Computing the posterior distribution of $\theta$ firstly implies to specify a prior $p(\theta)$ from which the initial particles $\{\theta_{0}^{i}\}_{1 \leq i \leq M_{\rm out}}$ will be drawn. For simplicity, we consider here uniform priors (as in \cite{bird2016bayesian,gontier2020identifiability}), although the algorithm readily extends to different choices of prior.

Similarly, initial samples for the hidden states $\{n_{0}^{i,j},k_{0}^{i,j}\}_{1 \leq j \leq M_{\rm in}}$ need to be drawn. For $i \in \{1,...,M_{\rm out}\}$, $j \in \{1,...,M_{\rm in}\}$, we define:

\begin{itemize}
    \item $n_0^{i,j} = N_i$ (i.e. all vesicles are supposed to be in the readily-releasable state at the beginning of the simulation);
    \item $k_0^{i,j} \sim {\rm Bin}(N_i,p_i)$ (consistently with Eq. \ref{eq:transition1}).
\end{itemize}

\textbf{Jittering step}: The parameters that we wish to infer are supposed to be constant. It is thus impossible to define dynamics of the form $p(\theta^i_{t+1}|\theta^i_{t})$ for the particles (as opposed to filtering problems aiming at inferring a dynamical hidden state, as for instance in \cite{kutschireiter2017nonlinear}). To avoid particle degeneracy, it is thus necessary to mutate particles using a jittering kernel $\kappa(\theta_{t-1}^{i})$. When particles take continuous values, a classical choice for the jittering kernel is to draw the next particle $\theta_{t}^{i}$ from a Gaussian distribution with mean $\theta_{t-1}^{i}$ and which variance $\iota$ is called the jittering width (see \cite{crisan2018nested} for a detailed discussion). In our implementation, the range of possible values for each parameter is discretized, so that each particle corresponds to a position on the grid of possible parameters values (same implementation as in \cite{bird2016bayesian}). The free parameter $\iota$ in our jittering kernel thus corresponds to the probability of moving by one bin:

\begin{equation}
   \theta_{t}^{i}= \kappa(\theta_{t-1}^{i})= 
\begin{cases}
    \theta_{t-1}^{i},& \text{with probability } 1-\iota\\
    \Tilde{\theta}_{t-1}^{i},              & \text{with probability } \iota
\end{cases}
\end{equation}

where $\Tilde{\theta}_{t-1}^{i}$ is one (randomly chosen) bin away from $\theta_{t-1}^{i}$.

\textbf{Propagation step}: Inner particles are redrawn based on $n_t^{i,j} \sim p(n_t^{i,j}|n_{t-1}^{i,j},k_{t-1}^{i,j},\theta_t^{i},x_t)$ (Equation \ref{eq:refill_prob}) and $k_t^{i,j} \sim p(k_t^{i,j}|n_{t}^{i,j},\theta_t^{i})$ (Equation \ref{eq:transition1}).

\textbf{Likelihood computation step}: $p(y_t|n_t^{i,j},k_t^{i,j},\theta_t^{i})$  is computed according to Equation \ref{eq:observation}.

\textbf{Resampling step}: Particles are resampled using the algorithm introduced in \cite{bentley1980generating}, which allows to draw a list of sorted numbers in a single step.

\begin{algorithm}[ht]
\SetAlgoLined

 \textbf{Input}: $\{\theta_{t-1}^{i}\}_{1 \leq i \leq M_{\rm out}}$, $\{n_{t-1}^{i,j},k_{t-1}^{i,j}\}_{1 \leq j \leq M_{\rm in}}$, $x_t$, $y_t$ \;
 
 {
 \For{$i$ in $1 \dots M_{\rm out}$}{
  \textbf{Jittering}: update the outer particles $\theta_t^{i} = \kappa(\theta_{t-1}^{i})$\;
  
  \For{$j$ in $1 \dots M_{\rm in}$}{
  
    \textbf{Propagation}: Draw  $n_t^{i,j} \sim p(n_t^{i,j}|n_{t-1}^{i,j},k_{t-1}^{i,j},\theta_t^{i},x_t)$ and $k_t^{i,j} \sim p(k_t^{i,j}|n_{t}^{i,j},\theta_t^{i})$\;
    
    \textbf{Likelihood}: compute $\Tilde{w}_t^{i,j} = p(y_t|n_t^{i,j},k_t^{i,j},\theta_t^{i})$ \;
    }
    \textbf{Normalization}: $\Tilde{w}_t^{i,j} \leftarrow \Tilde{w}_t^{i,j} / \sum_j \Tilde{w}_t^{i,j}$\;

  \textbf{Inner particles resampling}: resample $\{n_{t}^{i,j},k_{t}^{i,j}\}_{1 \leq j \leq M_{\rm in}}$ based on $\{\Tilde{w}_t^{i,j}\}_{1 \leq j \leq M_{\rm in}}$\;

  }
  Compute $w_t^{i} = \frac{1}{M_{\rm in}} \sum\limits_{j} \Tilde{w}_t^{i,j} $\;
  
  \textbf{Normalization}: $w_t^{i} \leftarrow w_t^{i} / \sum_i w_t^{i}$\;

  \textbf{Outer particles resampling}: resample $\{\theta_{t}^{i}\}_{1 \leq i \leq M_{\rm out}}$ and $\{n_{t}^{i,j},k_{t}^{i,j}\}_{1 \leq j \leq M_{\rm in}}$ based on $\{w_t^{i}\}_{1 \leq i \leq M_{\rm out}}$\;

 }
 \textbf{Output}: $\{\theta_t^{i}\}_{1 \leq i \leq M_{\rm out}}$, $\{n_t^{i,j},k_t^{i,j}\}_{1 \leq j \leq M_{\rm in}}$
 \caption{Particle filtering  for computing one step update of the posterior distribution of parameters}\label{algo:1}
\end{algorithm}


\begin{algorithm}[tb]
\SetAlgoLined

 set $\hat{\theta}_t = \frac{1}{M_{\rm out}} \sum_{i=1}^{M_{\rm out}} \theta_t^i$ (mean from the current posterior estimation)\;
 \textbf{Input}: $\mathcal{S}_{t+1}$ (set of candidates $x_{t+1}$)\;
 \For{$x_{t+1}$ in $\mathcal{S}_{t+1}$}{
  Compute $\mathbb{E}(Y_{t+1}|x_{1:t+1},\hat{\theta}_t)$ using Eq. \ref{eq:mean}\;
  Compute $H_{x_{t+1}} (\Theta|h_t,Y_{t+1}=\mathbb{E}(Y_{t+1}|x_{1:t+1},\hat{\theta}_t))$ using Algorithm \ref{algo:1}\;
 }
 $x_{t+1}^* = \argmin_{x_{t+1} \in \mathcal{S}_{t+1}} H_{x_{t+1}} (\Theta|h_t,Y_{t+1}=\mathbb{E}(Y_{t+1}|x_{1:t+1},\hat{\theta}_t))$
 \caption{Computation of the optimal next stimulation time for synaptic characterization}\label{algo:2}
\end{algorithm}

Algorithm \ref{algo:2} is slightly modified in Figure \ref{fig:fig3} (a) for the "ESB-BAL (exact)" simulations, in which Eq. \ref{eq:no_simplification} is computed using MC samples instead of the point-based simplifications explained in Eqs. \ref{eq:simplification1} and \ref{eq:simplification2}. Samples to compute the expectation over $\theta$ are drawn from the current posterior distribution $p(\theta|h_t)$, i.e. by random sampling from the pool of particles $\{\theta^i_t \}_{i \in \{1,\dots,M_{out} \}}$. For each of these samples, and for each candidate next input $x_{t+1}$ in $\mathcal{S}_{t+1}$, samples used to compute the expectation over $y_{t+1}$ are drawn by randomly sampling $n_{t+1}$, $k_{t+1}$, and $y_{t+1}$ (using respectively Eqs \ref{eq:refill_prob}, \ref{eq:transition1}, and \ref{eq:observation}) from the ground-truth values of the hidden states $n_t$ and $k_t$.

\subsection{Mean-field approximation of vesicle dynamics}\label{sec:mean_field}

Our synapse model, as defined by Eq. \ref{eq:p_e} to \ref{eq:refill_prob}, is a Hidden Markov Model with observations $y_t$ and hidden states $z_t = (n_t,k_t)$. 
The predictive distribution $p(y_{t+1}|h_t,x_{t+1},\theta)$ used in Eq. \ref{eq:mutual_info} can be computed using the Baum-Welch algorithm: however, the algorithmic complexity of this forward-backward procedure, which scales with $N^4$, makes it impractical for closed-loop applications. 
Here, we suggest that computation can be massively simplified by using a mean-field approximation of vesicle dynamics: the analytical mean and variance of hidden and observed variables can be computed using recursive formul\ae.

Let $r_t \in [0,1]$ denote the average fraction of release-competent vesicles at the moment of spike $t$. Its values, given $\theta = [N,p,q,\sigma,\tau_D]$ and $x_{1:t}$, can be iteratively computed (see \cite{barri2016quantifying}, Eq. (7)) from the equations of the Tsodyks-Markram model \citep{tsodyks1998neural}:

\begin{equation}
r_t = 1-(1-(1-p)r_{t-1})\exp \left (-\frac{x_t}{\tau_D} \right )
\end{equation}

with $r_1 = 1$. It follows that the expected value of the EPSC after spike $t$ is 
\begin{equation} \label{eq:mean}
    \mathbb{E}(Y_t|x_{1:t},\theta) = r_t N p q
    \end{equation}

One can note that the variance of the number of available vesicles $n_t$ conditioned on the history of previous activations $x_{1:t}$ and on the parameter values $\theta$  can be computed similarly using the law of total variance:

\begin{equation}
\mathrm{Var}(n_{t}|x_{1:t},\theta) = \mathbb{E}(\mathrm{Var}(n_{t}|n_{t-1},k_{t-1},x_{1:t},\theta)) + \mathrm{Var}(\mathbb{E}(n_{t}|n_{t-1},k_{t-1},x_{1:t},\theta))
\end{equation}

Since $n_{t} = n_{t-1} -k_{t-1} + v_t$
with
$v_t \sim \mathrm{Bin}({N-n_{t-1}+k_{t-1}, \pi(x_t)})$ (see Eq. \ref{eq:refill_prob}), it follows that 

\begin{equation}
\mathrm{Var}(n_{t}|x_{1:t},\theta) = \pi(x_t)(1-\pi(x_t))N(1-r_{t-1}+p r_{t-1}) + (1-\pi(x_t))^2 \mathrm{Var}(n_{t-1}-k_{t-1}|x_{1:t-1},\theta)
\end{equation}

Finally, by noting that $(n_t - k_t) | n_t \sim \mathrm{Bin}({n_{t}, 1-p})$ and using again the law of total variance to compute

\begin{equation}
\mathrm{Var}(n_{t-1}-k_{t-1}|x_{1:t-1},\theta) = \mathbb{E}(\mathrm{Var}(n_{t-1} - k_{t-1} | n_{t-1},x_{1:t-1},\theta)) + \mathrm{Var}(\mathbb{E}(n_{t-1} - k_{t-1} | n_{t-1},x_{1:t-1},\theta)) 
\end{equation}

we obtain

\begin{equation} \label{eq:variance}
\mathrm{Var}(Y_t|x_{1:t},\theta) = \sigma^2 + q^2 (Nr_t p(1-p) + \mathrm{Var}(n_{t}|x_{1:t-1},\theta)p^2)
\end{equation}

\subsection{Third setting: batch optimization and application to neural recordings}\label{sec:appendix_third_setting}

Each candidate batch of $n$ stimulation times in $\mathcal{S}_{t+1:t+n}$ (Figure \ref{fig:fig5} (b)) is described by 3 parameters:
\begin{itemize}
    \item $m < n$: the number of tetanic stimulations [-];
    \item $f$: the frequency of the tetanic stimulations [Hz];
    \item $x^{\rm last}$: the time interval before the final recovery spike [s].
\end{itemize}

A train of $n$ stimulations is thus composed of $m$ tetanic stimulations at a frequency $f$, followed by $n-m$ recovery spikes with increasing inter-spike intervals $\frac{x^{\rm last}}{n-m},\frac{x^{\rm last}}{n-m-1},\dots,\frac{x^{\rm last}}{2},x^{\rm last}$. The following values were used during experiments (Figure \ref{fig:fig5} (a)): $n = 26$, $m \in [5,10,15,20]$, $f \in [25Hz, 50Hz, 100Hz, 200Hz]$, $x^{\rm last} \in [0.1s,0.5s,1.0s,2.0s]$.

\begin{algorithm}[ht]
\SetAlgoLined

 \textbf{Input}: $\mathcal{S}_{t+1:t+n}$ (set of candidates $x_{t+1:t+n}$)\;
 \For{$x_{t+1:t+n}$ in $\mathcal{S}_{t+1:t+n}$}{
  Compute $H_{x_{t+1:t+n}} (\Theta|h_t,Y_{t+1:t+n})$ using Algorithm \ref{algo:1}\;
 }
 $x_{t+1:t+n}^* = \argmin_{x_{t+1:t+n} \in \mathcal{S}_{t+1:t+n}} H_{x_{t+1:t+n}} (\Theta|h_t,Y_{t+1:t+n})$
 \caption{Computation of the optimal next batch of ISIs}\label{algo:3}
\end{algorithm}

\subsection{Electrophysiology recordings}

Animals were treated following national and institutional guidelines. The Cantonal Veterinary Office of Zurich approved all experiments (authorization no. ZH009/2020). Experiments were performed in male and female 1–2-month-old C57BL/6J mice (Janvier Labs, France). Animals were housed in groups of 3–5 in standard cages on a 12h-light/12h-dark cycle with food and water ad libitum. Mice were sacrificed by rapid decapitation after isoflurane anesthesia. The cerebellar vermis was removed quickly and mounted in a chamber filled with cooled extracellular solution. 300-µm thick parasagittal slices were cut using a Leica VT1200S vibratome (Leica Microsystems, Germany), transferred to an incubation chamber at 35 $^{\circ}$C for 30 minutes, and then stored at room temperature until experiments.

The extracellular solution (artificial cerebrospinal fluid, ACSF) for slice cutting and storage contained (in mM): 125 NaCl, 25 NaHCO3, 20 D-glucose, 2.5 KCl, 2 CaCl2, 1.25 NaH2PO4, 1 MgCl2, bubbled with 95\% O2 and 5\% CO2. Slices were visualized using an upright microscope with a 60×, 1 NA water-immersion objective, infrared optics, and differential interference contrast (Scientifica, UK). The recording chamber was continuously perfused with ACSF supplemented with 10 µM D-APV, 10 µM bicuculline, and 1 µM strychnine. Experiments were performed at room temperature (21–25 $^{\circ}$C). Patch pipettes (open-tip resistances of 3–8 M$\Omega$) were filled with solution containing (in mM): 150 K-D-gluconate, 10 NaCl, 10 HEPES, 3 MgATP, 0.3 NaGTP, 0.05 ethyleneglycol-bis(2-aminoethylether)-N,N,N',N'-tetraacetic acid (EGTA), pH adjusted to 7.3 using KOH. 

Voltage-clamp recordings were done using a HEKA EPC10 amplifier controlled via Patchmaster software (HEKA Elektronik GmbH, Germany) essentially as described in \citep{kita2021glua4}. Voltages were corrected for a liquid junction potential of +13 mV. Extracellular mossy fiber stimulation was performed using square voltage pulses (duration, 150 µs) generated by a stimulus isolation unit (ISO-STIM 01B, NPI) and applied through an ACSF-filled pipette. The pipette was moved over the slice surface close to the postsynaptic cell while applying voltage pulses until excitatory postsynaptic currents (EPSCs) could be evoked reliably. Care was taken to stimulate single mossy fiber inputs. EPSCs were recorded at a holding potential of –80 mV; data were low-pass filtered at 2.9 kHz and digitized at 20–50 kHz. Train stimulation protocols comprised bouts of 20 or 100 MF stimulations at 100 Hz, followed by single pulses to monitor recovery from short-term depression (intervals: 25 ms, 50 ms, 100 ms, 300 ms, 1 s, 3 s). The interval between subsequent train recordings was at least 30 s. For OED experiments, custom protocols were generated online as file templates for use with Patchmaster. EPSCs were quantified as peak amplitudes from a 300-µs baseline before onset.

To facilitate the definition of the range of possible values for parameters $q$ and $\sigma$ (and especially to avoid running an experiment with too narrow ranges), recorded EPSC amplitudes were normalized by dividing them by their maximum value. Analyses are thus performed by assuming $q \in [0,1]$ and  $\sigma \in [0,1]$. Posterior plots in Figure \ref{fig:fig5} (a) were then multiplied by the maximum amplitudes observed in each dataset.

\section{Acknowledgments}

The work presented in this paper was supported by the Swiss National Science Foundation 
under grant number 31003A\_175644 entitled "Bayesian Synapse". 

Calculations were performed on UBELIX (\url{http://www.id.unibe.ch/hpc}), the HPC cluster at the University of Bern. The CUDA.jl package \citep{besard2018juliagpu,besard2019prototyping} is licensed under the MIT "Expat" License (\url{https://github.com/JuliaGPU/CUDA.jl/blob/master/LICENSE.md}). 

We thank Ehsan Abedi, Jakob Jordan, and Anna Kutschireiter for the fruitful discussions.

JULIA files are available in the following package: \url{https://github.com/Theoretical-Neuroscience-Group/BinomialSynapses.jl}

\section{Competing interests}

The authors declare that the research was conducted in the absence of any commercial or financial relationships that could be construed as a potential conflict of interest.

\bibliography{elife-sample}


\appendix

\clearpage

\section{Appendix}

\begin{figure}[ht]
  \centering
  \includegraphics[width=10cm]{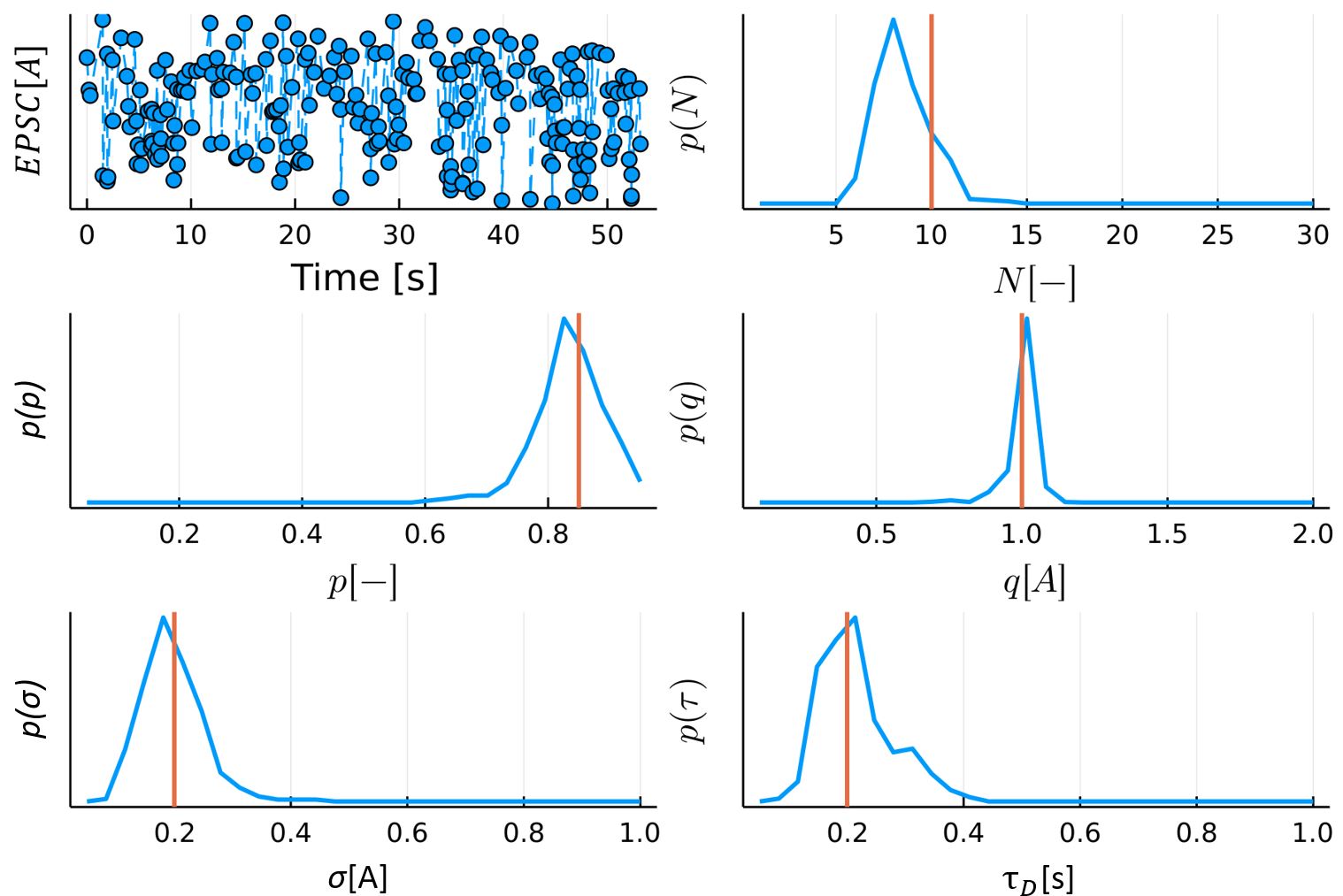} 
  \caption{Examples of posteriors obtained using the filter (Algorithm \ref{algo:1}).
   Upper left panel: train of synthetic EPSCs generated from the model described in Section \nameref{sec:system2}. Other panels: posterior distributions of the parameters after 230 stimulations. Ground-truth values used to generate the EPSCs are displayed as red vertical lines.}\label{fig:posterior}
\end{figure}

\begin{figure}[ht]
  \centering
  \includegraphics[width=10cm]{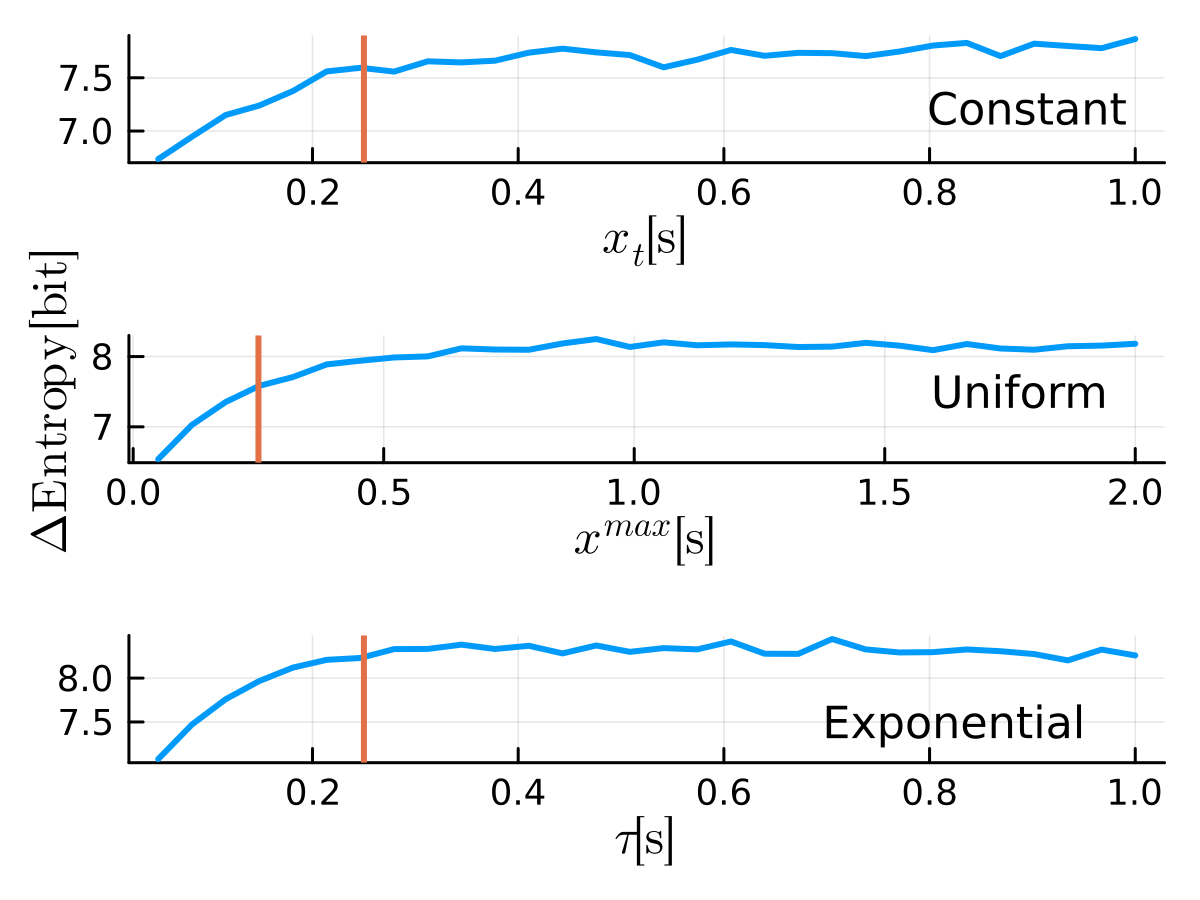} 
  \caption{Average final entropy decrease (i.e. information gain) after 200 observations using the \textit{Constant} (top), \textit{Uniform} (middle), or \textit{Exponential} (bottom) protocol, for different values of their hyperparameters. Ground truth parameters used are $N^* = 7$, $p^* = 0.6$, $q^* = 1$ A, $\sigma^* = 0.2$ A, and $\tau_D^* = 0.25$s \citep{bird2016bayesian}. Vertical red lines indicate the ground truth value $\tau_D^* = 0.25$s used for simulations.
Optimal values for $x_t$, $x^{\rm max}$, and $\tau$ are used in Figure \ref{fig:fig3}.}\label{fig:results_optim2}
\end{figure}

\begin{figure}
\centering
\begin{minipage}[c][][c]{.4\textwidth}
  \vspace*{\fill}
  \centering
  \caption*{(a)}
  \includegraphics[width=1\textwidth]{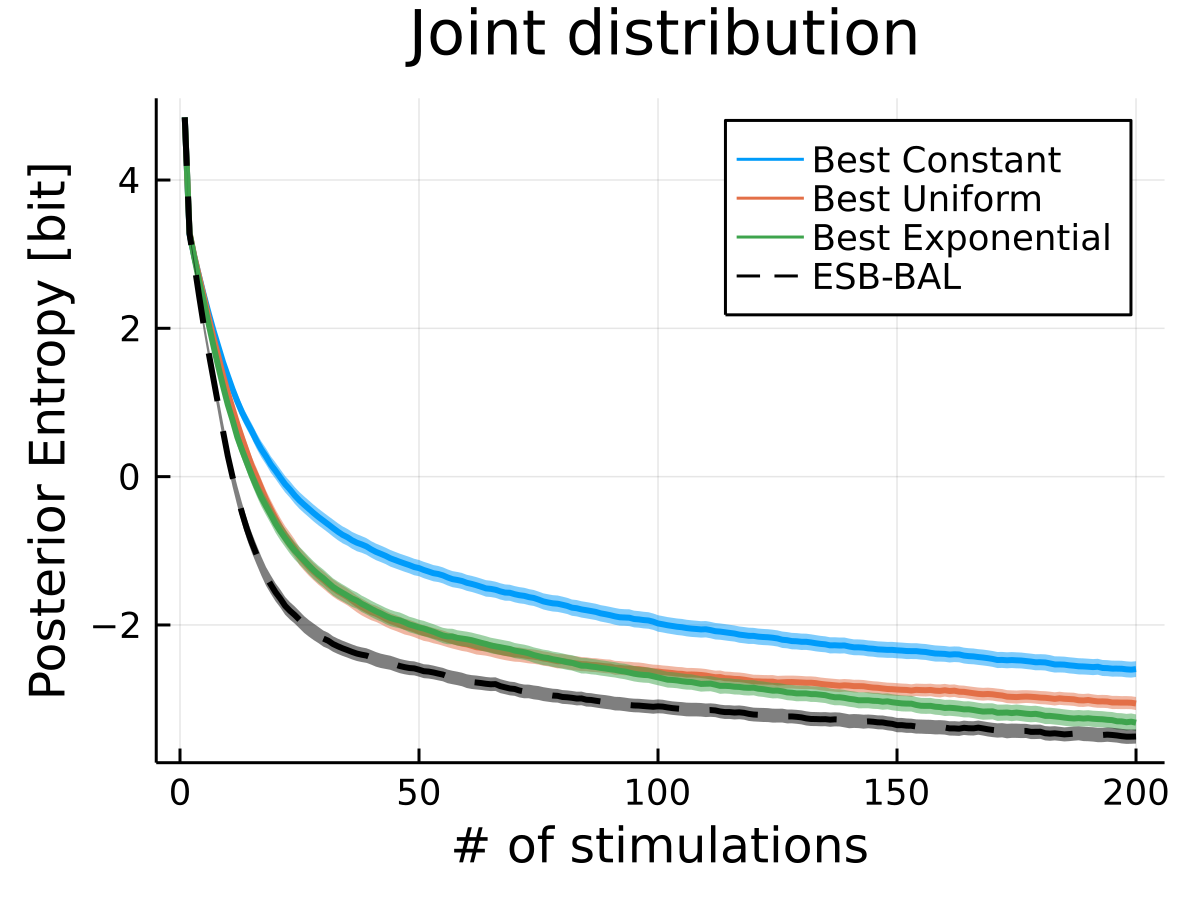}
  
\end{minipage}%
\begin{minipage}[c][][c]{.4\textwidth}
  \vspace*{\fill}
  \centering
  \caption*{(b)}
  \includegraphics[width=1\textwidth]{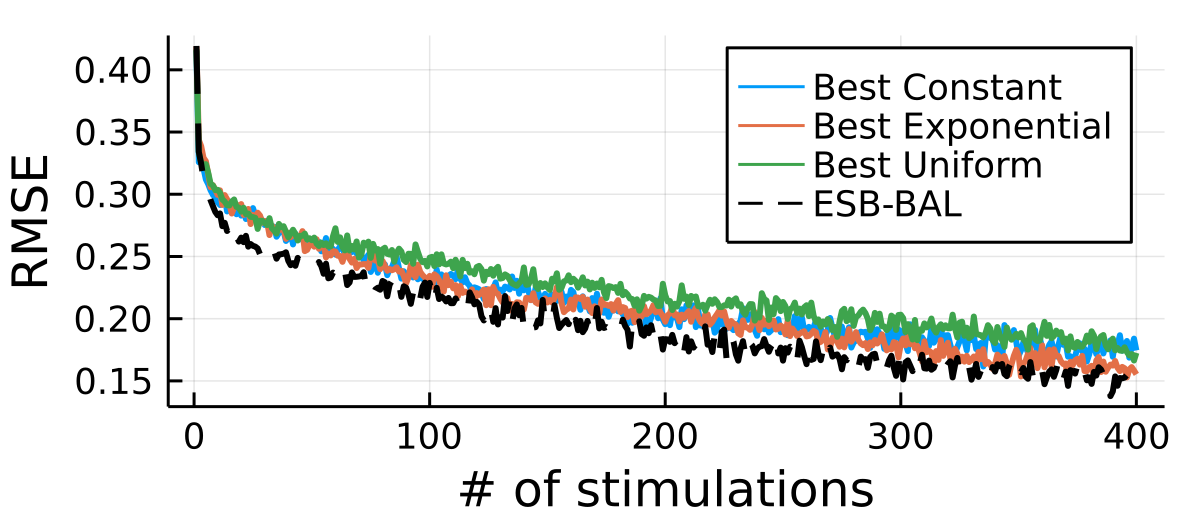}
  
  \par\vfill
  \caption*{(c)}
  \includegraphics[width=1\textwidth]{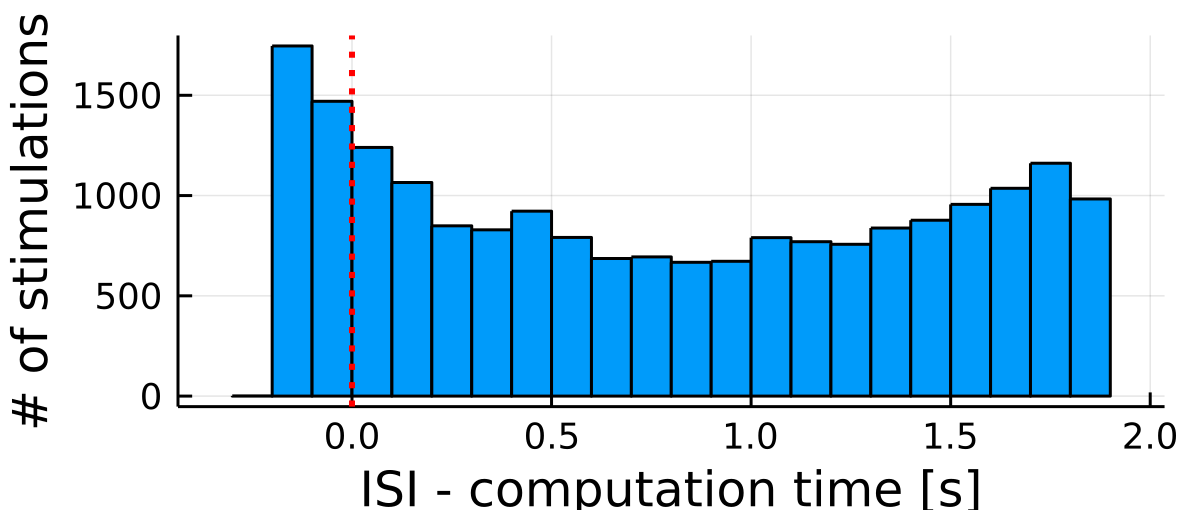}
  
\end{minipage}

\caption{Same setting as in Figure \ref{fig:fig3} but for  ground truth parameters $N^* = 10$, $p^* = 0.85$, $q^* = 1$ A, $\sigma^* = 0.2$ A, and $\tau_D^* = 0.2$s.}
\label{fig:fig8}
\end{figure}

\begin{figure}
\centering
\begin{minipage}[c][][c]{.4\textwidth}
  \vspace*{\fill}
  \centering
  \caption*{(a)}
  \includegraphics[width=1\textwidth]{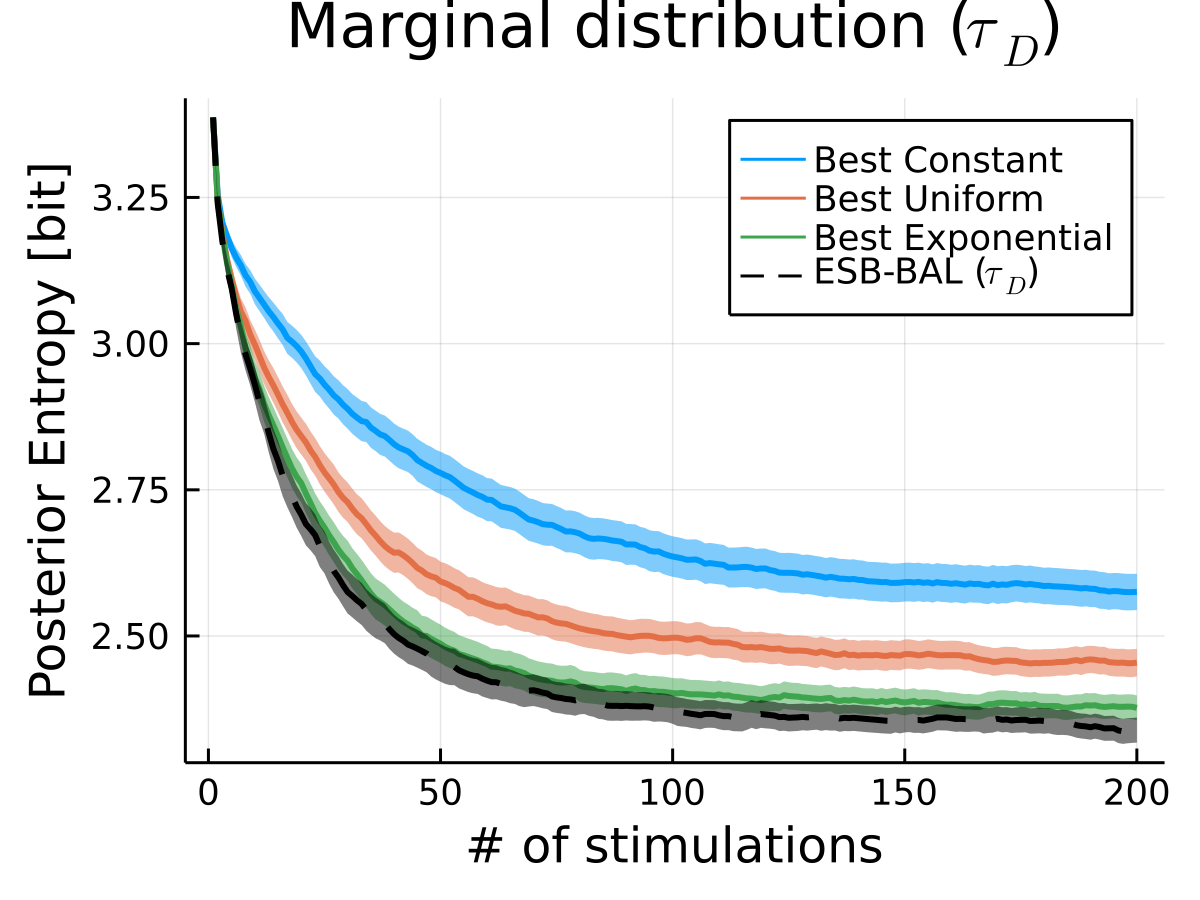}
  
\end{minipage}%
\begin{minipage}[c][][c]{.4\textwidth}
  \vspace*{\fill}
  \centering
  \caption*{(b)}
  \includegraphics[width=1\textwidth]{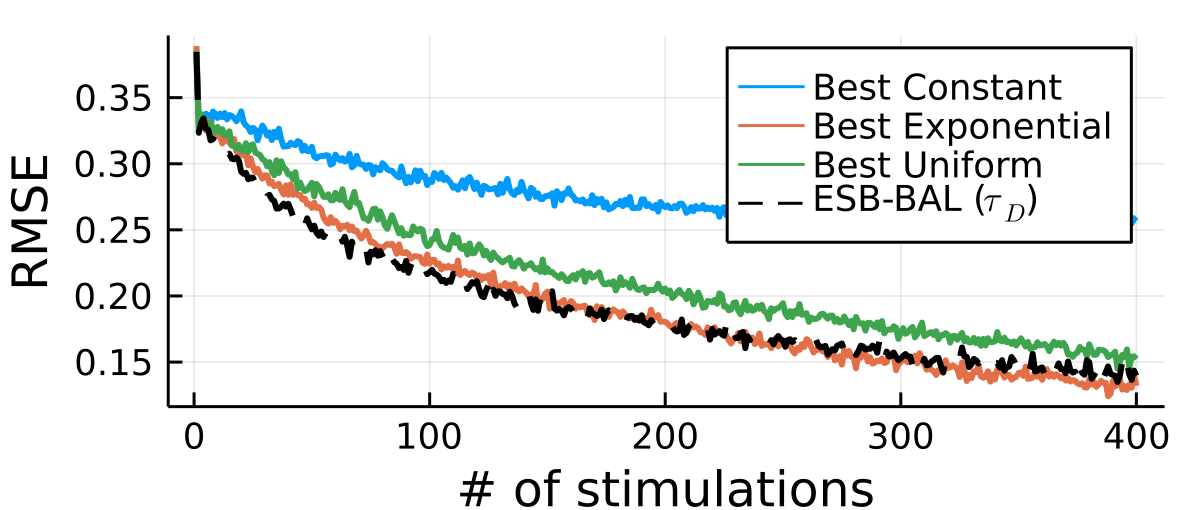}
  
  \label{fig:test2}\par\vfill
  \caption*{(c)}
  \includegraphics[width=1\textwidth]{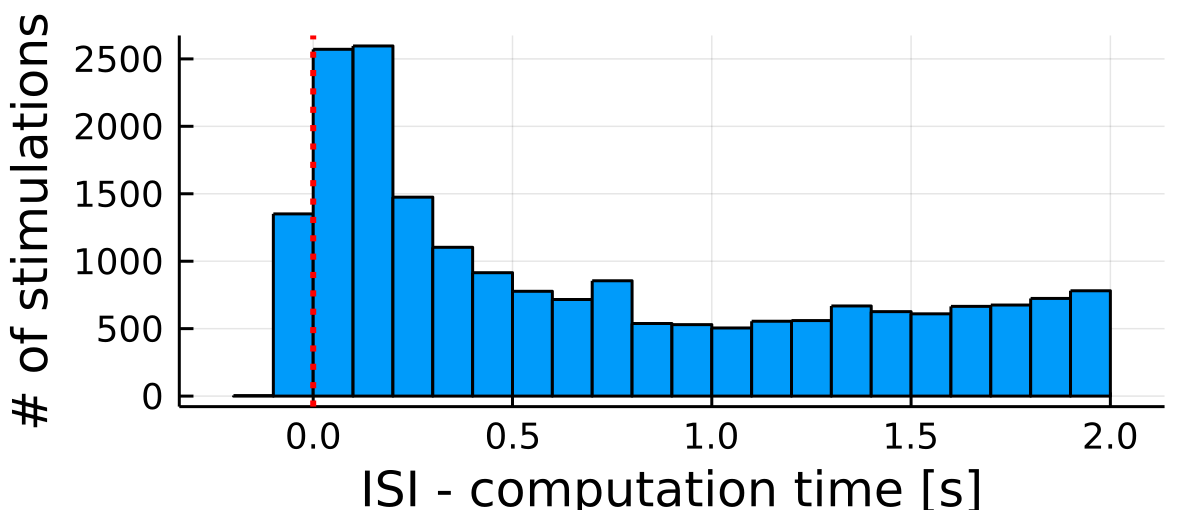}
  
\end{minipage}

\caption{Same setting as in Figure \ref{fig:fig3} but when optimizing solely for the marginal posterior distribution of $\tau_D$.}
\label{fig:fig7}
\end{figure}

\begin{figure}[ht]
  \centering
  \includegraphics[width=10cm]{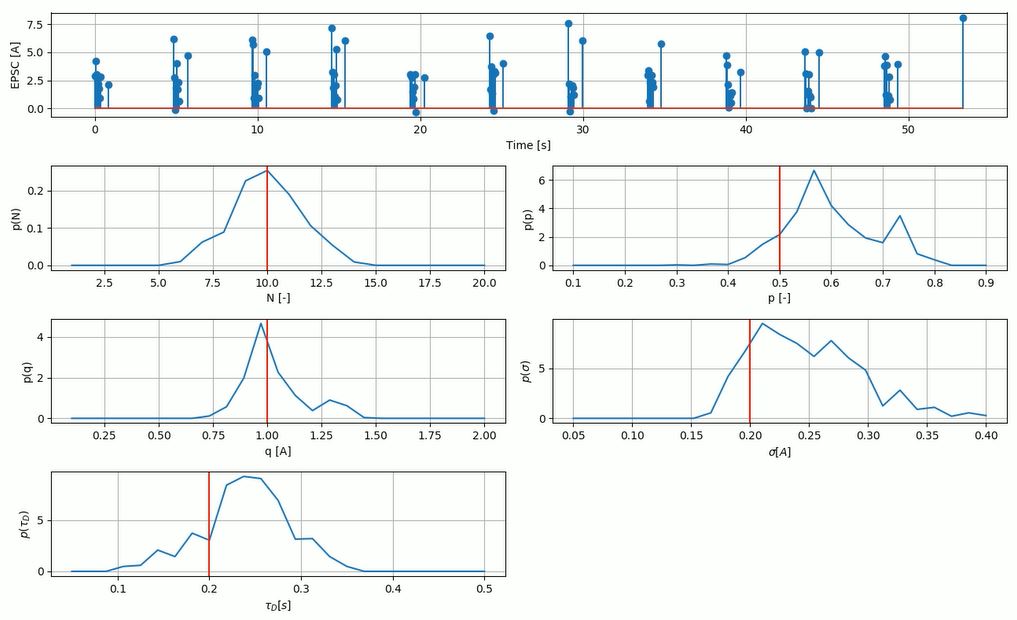} 
  \caption[Illustration of the Nested Particle Filter (NPF)]{\textbf{Illustration of the Nested Particle Filter (NPF)}. The NPF \cite{crisan2018nested} is a non-linear particle filtering algorithm used to infer parameters of HMMs. It is asymptotically exact and purely recursive, and thus allows to directly estimate the distribution of parameters as recordings are acquired. 
The NPF relies on two nested layers of particles to approximate the posterior distributions of both the static parameters of the model and of its hidden states. 
A first outer filter with $M_{\rm out}$ particles is used to compute the posterior distribution of parameters $p(\theta|y)$, and for each of these particles, an inner filter with $M_{\rm in}$ particles is used to estimate the corresponding hidden states (so that the total number of particles in the system is $M_{\rm out} \times M_{\rm in}$).
After each new observation, these particles are resampled based on their respective likelihoods. Its implementation for synaptic characterization is detailed in Section \nameref{sec:filter}. Link to the video: \url{https://youtu.be/OPGEyayhxJI}. Upper panel: train of synthetic EPSCs generated from a model of synapse with short-term depression. Lower panels: posterior distributions of the parameters computed online using the nested particle filter. Ground-truth values used to generate the EPSCs are displayed as red vertical lines.}\label{fig:illus_NPF}
\end{figure}

\end{document}